%% file: o4a_cw_all-sky_binary.tex
\documentclass[%
 reprint,
 superscriptaddress,
nofootinbib,
 amsmath,amssymb,
 aps,
]{revtex4-2}

\usepackage{orcidlink}
\usepackage{xcolor}
\usepackage{booktabs}
\usepackage{graphicx}
\usepackage{dcolumn}
\usepackage{bm}
\usepackage[mathlines]{lineno}
\usepackage{sidecap}

\newtoggle{fullauthorlist}

\newtoggle{endauthorlist}
\toggletrue{endauthorlist}

\newcommand{\Tsft}{T_{\text{SFT}}}

\begin{document}


\title{All-sky search for continuous gravitational-wave signals from unknown neutron stars in binary systems in the first part of the fourth LIGO-Virgo-KAGRA observing run}

\iftoggle{endauthorlist}{
 %
 %
 \let\mymaketitle\maketitle
 \let\myauthor\author
 \let\myaffiliation\affiliation
 \author{The LIGO Scientific Collaboration, the Virgo Collaboration and the KAGRA Collaboration}
}{
 %
 %
 \iftoggle{fullauthorlist}{
  \input{LSC-Virgo-KAGRA-Authors-Feb-2025-prd}
 }{
 \author{The LIGO Scientific Collaboration and the Virgo Collaboration and the KAGRA Collaboration}
}
}

\date{\today}

\begin{abstract}
    We present the results of a blind all-sky search for continuous gravitational-wave signals from neutron stars in binary systems using data from the first part of the fourth observing run (O4a) using LIGO detectors data. Rapidly rotating, non-axisymmetric neutron stars are expected to emit continuous gravitational waves, whose detection would significantly improve our understanding of the galactic neutron star population and matter under extreme conditions, while also providing valuable tests of general relativity. Neutron stars in binary systems likely constitute a substantial fraction of the unobserved galactic population and, due to potential mass accretion, may emit stronger gravitational-wave signals than their isolated counterparts. This search targets signals from neutron stars with frequencies in the $100-350$~Hz range, with orbital periods between 7 and 15~days and projected semi-major axes between 5 and 15 light-seconds. The analysis employs the GPU-accelerated \texttt{fasttracks} pipeline. No credible astrophysical signals were identified, and, in the absence of a detection, we report search sensitivity estimates on the population of neutron stars in binary systems in the Milky Way.
\end{abstract}

\maketitle

\section{\label{sec:introduction} Introduction}

Continuous gravitational waves (CWs) are long-lasting, nearly monochromatic signals. One of the most promising sources of CWs are rapidly rotating neutron stars (NSs) with non-axisymmetric mass distributions~\cite{Riles:2022wwz, Sieniawska:2019hmd}. These gravitational waves (GWs) arise from the star's time-varying quadrupole moment and can be generated through several mechanisms, which in turn provide a unique probe of the elastic, magnetic, thermal, and superfluid properties of NSs, both in isolated and binary systems~\cite{Lasky:2015uia, Sieniawska:2019hmd}.

Non-axisymmetric distortions (“mountains”) sustained either by elastic stresses in the crust or by strong magnetic fields can lead to the emission of CWs. Accreting systems in particular may generate sizable deformations~\cite{Ushomirsky:2000ax, Horowitz:2009vm, Bildsten:1998ey}. Furthermore, non-axisymmetric instabilities can drive oscillation modes, such as r-modes, that may emit GWs, although their strength is strongly dependent on damping mechanisms and uncertain saturation amplitudes~\cite{Andersson:1997xt, Andersson:2002ch}. Finally, free precession caused by a misalignment between the rotation and symmetry axes can lead to quasi-periodic emission~\cite{VanDenBroeck:2004wj}.

The maximum possible ellipticity of a NS depends strongly on its interior composition and the poorly constrained equation of state, with estimates ranging from $\sim10^{-6}$ for hadronic matter to $\sim10^{-4}$ in scenarios allowing for more exotic physics such as crystalline phases~\cite{Morales:2022wxs, Ushomirsky:2000ax}. Additional factors such as general relativistic effects and strong magnetic fields can further influence the maximum sustainable deformation~\cite{Soldateschi:2021cxq, Colaiuda:2007br}. By comparing the observational spindown limits on the ellipticity with these theoretical expectations, current CW searches probe astrophysically relevant regions of parameter space, and can provide constraints on the physics of dense matter~\cite{Woan:2018tey}.

NSs are among the densest known objects in the universe, with core densities of the order of nuclear values~\cite{Haskell:2023yrv}. This makes them unique laboratories for probing the composition and behavior of matter under conditions not achievable in terrestrial laboratories, as well as studying potential deviations from general relativity, e.g. by looking for wave polarizations different from general relativity \cite{Lasky:2015uia}.

Highly magnetized rotating NSs can emit beams of electromagnetic radiation that are detected as pulses, in which case they are known as pulsars. A significant portion of the known pulsar population resides in binary systems \cite{Manchester:2004bp}, where mass accretion from a binary companion may increase the quadrupolar deformation of the NS, causing it to emit stronger CW signals than their isolated counterparts \cite{Bildsten:1998ey, 2014MNRAS.445..385C, 2018ApJ...857...38H, 2021MNRAS.507..116G, Gittins:2024zbg, Rossetto:2023btr}. Moreover, a large fraction of the known millisecond pulsars are in binary systems, spinning at frequencies that are in the sensitive band of ground-based GW detectors. Therefore, NS in binary systems are promising targets for CW detection in all-sky searches with data from current ground-based interferometers (IFOs).

Although numerous CW searches have been conducted to date, no detection has yet been achieved (see \cite{Tenorio:2021wmz, Piccinni:2022vsd, Riles:2022wwz, Wette:2023dom} for recent reviews, and \cite{LIGOScientific:2025kei, LIGOScientific:2022enz, KAGRA:2022dqk, KAGRA:2022dwb, LIGOScientific:2021quq, LIGOScientific:2021inr, LIGOScientific:2021rnv, LIGOScientific:2021hvc, LIGOScientific:2021ozr, KAGRA:2021una, LIGOScientific:2021mwx, LIGOScientific:2021yby, LIGOScientific:2020lkw, Ming:2024dug, Steltner:2023cfk, 2025arXiv250816423M, 2025arXiv250820073M} for results from the O3 and O4 LIGO-Virgo-KAGRA observing runs~\cite{LIGOScientific:2014pky, VIRGO:2014yos, KAGRA:2018plz}). In comparison with signals from compact binary mergers, from which all GW detections to date have originated~\cite{LIGOScientific:2018mvr, LIGOScientific:2020ibl, KAGRA:2021vkt, 2020PhRvD.102f2003B, Capote:2024rmo, LIGOScientific:2025slb}, CW amplitudes are expected to be several orders of magnitude smaller, making them challenging to detect. Moreover, CW signals are subject to significant frequency modulation caused by the Earth's rotation and orbital motion \cite{Jaranowski:1998qm, Cutler:2005hc, Wette:2014tca}. As a result, all-sky searches must cover wide ranges of frequencies, sky positions, and in the case of binaries, orbital parameters \cite{Wette:2023dom}. These factors enlarge the search parameter space and require vast numbers of signal templates, up to more than $10^{16}$ templates for all-sky searches such as the one presented here.

To make CW searches computationally feasible, semi-coherent methods are commonly employed~\cite{PhysRevD.57.2101,Krishnan:2004sv, PhysRevD.72.042004, PhysRevD.85.084010}. These approaches divide the data into shorter coherent segments, combine them incoherently, and follow up the most significant candidates with longer coherence times. This strategy reduces the computational load by allowing us to cover the parameter space less densely with templates and by improving robustness to small mismatches with the signal model, such as unmodeled astrophysical effects~\cite{PhysRevD.97.043016}.

All-sky searches aim to detect CW emission from unknown NSs within our galaxy, that is, those that have not been observed using electromagnetic telescopes neither as pulsars nor as non-pulsating sources. Since only a small fraction of the estimated NS population has been observed as pulsars~\cite{Rajwade:2018xxf, Reed:2021scb}, conducting this type of searches offers a valuable opportunity to find these undetected NSs.

To date, several pipelines have been developed and implemented for all-sky CW searches targeting unknown NSs in binary systems. These include the \texttt{TwoSpect} pipeline \cite{Goetz:2011bd}, and the \texttt{BinarySkyHough} pipeline~\cite{Covas:2019jqa}, which has been employed in searches using data from the second~\cite{Covas:2020nwy} and third observing runs~\cite{LIGOScientific:2020qhb}. More recently, the \texttt{BinarySkyHou$\mathcal{F}$} pipeline~\cite{Covas:2022mqo, Covas:2022rfg, Covas:2024nzs} has been introduced and the Falcon pipeline is being extended to include binary systems~\cite{2025arXiv250311503D}. \texttt{BinarySkyHough} is based on the SkyHough~\cite{Krishnan:2004sv} method, which uses the Hough transform~\cite{Hough1962} to search for CWs semi-coherently with the Hough number count detection statistic (see~\cite{Krishnan:2004sv, 2014PhRvD..90d2002A, PhysRevD.98.102004, PhysRevD.99.104067, Covas:2019jqa} for other works that use the Hough transform for searches of long-duration GW signals). Each of these pipelines makes different trade-offs between sensitivity and parameter space coverage, sacrificing some sensitivity in order to scan wider or more complex parameter spaces.

We present an all-sky search for CWs from unknown NSs in binary systems, carried out with the newly developed \texttt{fasttracks} pipeline \cite{Tenorio:2024jgc}. \texttt{fasttracks} is a massively parallel Python engine built on \texttt{JAX}~\cite{jax2018github}, designed to evaluate detection statistics for generic CW signals on Short Fourier Transform (SFT)-like data \cite{Allen:T040164}. This includes the detection statistics used in the first stage of this search, which are summed weighted normalized power from the SFTs and summed weighted Hough number count. \texttt{fasttracks} can exploit Graphics Processing Units (GPUs) to accelerate the calculation of detection statistics across large amount templates needed for all-sky searches. Since this template evaluation is by far the most computationally expensive step, GPU acceleration makes large-scale searches feasible.

In addition, we employ a novel parameter-space partitioning strategy originally introduced in \cite{Tenorio:2024jgc}, which divides the high-dimensional search space into smaller regions, or ``boxes''. Each box is searched independently, producing its own output. This approach has two main advantages: different boxes can be computed in parallel in different GPUs and it avoids the need for clustering procedures in post-processing. Thanks to these improvements, \texttt{fasttracks} achieves higher computational efficiency compared to previous searches such as \texttt{BinarySkyHough}~\cite{LIGOScientific:2020qhb} and streamlines analysis with the new parameter-space partitioning strategy.

The remainder of this paper is organized as follows. in Sec.~\ref{sec:sigmal_model}, we outline the CW signal model. Sec.~\ref{sec:data_used} describes the usage of data from the early fourth observing run (O4a). Sec.~\ref{sec:search_pipeline} presents the details of the search pipeline. In Sec.~\ref{sec:post_processing}, we describe the line veto procedure. Sec.~\ref{sec:sensitivity} contains the sensitivity depth estimates of the search. Sec.~\ref{sec:follow_up} discusses the most significant outliers identified, and explains how they were ruled out as non-astrophysical. The search results are discussed in Sec.~\ref{sec:results}. Finally, our conclusions are summarized in Sec.~\ref{sec:conclusion}.

\section{\label{sec:sigmal_model} Signal model}

A NS exhibiting a deviation from perfect axial symmetry with respect to its rotation axis is expected to emit CWs at twice its rotational frequency. This emission arises from the time-varying mass quadrupole moment associated with the star's asymmetry. The resulting GW strain amplitude $h_0$, which can be measured by interferometric detectors, depends on the physical properties of the source and can be expressed as \cite{Riles:2022wwz}
\begin{equation}
    h_0 = \frac{4 \pi^2 G}{c^4} \frac{I_{\textrm{zz}} \epsilon}{d} f_0^2,
    \label{eq:h0}
\end{equation}
where $G$ is the gravitational constant, $c$ is the speed of light, $I_{\textrm{zz}}$ is the principal moment of inertia around the rotation axis (assumed to be the z-axis), $d$ is the distance from the source to the detector, and $f_0$ is the intrinsic GW frequency. The equatorial ellipticity of the star, $\epsilon$, can be related to the mass quadrupole moment $Q_{22}$ via~\cite{Ushomirsky:2000ax}
\begin{equation}
    \epsilon = \sqrt{\frac{8\pi}{15}} \frac{\left|Q_{22}\right|}{I_{\textrm{zz}}}.
\end{equation}

Due to the motion of Earth-based detectors relative to the Solar System Barycenter (SSB) and the motion of the NS around the binary system barycenter, the observed CW signal is subject to Doppler modulation from both effects. As a result, for the circular, non-relativistic binary orbit of the unknown searched NSs, the GW frequency measured at the detector is time-dependent and can be expressed as \cite{Covas:2019jqa, Tenorio:2024jgc}
\begin{equation}
    f(t; \lambda) = f_0 \left[ 1 + \frac{\vec{v}(t)}{c} \cdot \hat{n} - a_p \Omega \cos(\Omega t - \phi_{\mathrm{b}}) \right],
    \label{eq:track}
\end{equation}
where $\vec{v}\left(t\right)$ is the detector's velocity vector relative to the SSB, $\hat{n}$ is a vector that points from the SSB to the sky location of the source, parametrized by $\alpha$, the right ascension and $\delta$, the declination. $f_0$ is the GW frequency emitted by the source NS, $a_p$ is the projected semi-major axis of the binary orbit in light-seconds, $\Omega$ is the angular orbital frequency of the source, and $\phi_{\mathrm{b}}$ is the initial orbital phase. The last three parameters describe the orbital motion of the NS in the binary system.

The resulting frequency evolution, or \textit{track}, traced by the signal in time-frequency space is the target of the search. It is parameterized by six quantities: $\lambda=\left\{ f_{0},\hat{n},a_{p},\Omega,\phi_{\mathrm{b}}\right\} $, where $\hat{n}$ is composed of $\alpha$ and $\delta$.

The model described above assumes a circular, non-relativistic binary orbit. However, the search remains sensitive to signals from NSs in slightly eccentric systems, with an upper limit on the allowed eccentricity given by \cite{Covas:2019jqa}
\begin{equation}
    e \ge \frac{1}{2 T_{\text{SFT}} f_0 a_p \Omega},
    \label{eq:eccentricty}
\end{equation}
where $T_\textrm{SFT}$ is the duration of each of the SFTs used in the analysis.

We assume that the NS does not undergo any glitches during the observing time and that spin wandering (stochastic fluctuations in the star's rotation frequency) can be neglected~\cite{Carlin:2025sxm}. The typical glitch rate for known pulsars is of order one per year or less~\cite{Wang:2000nw}. Estimates of spin wandering for accreting NSs suggest much smaller drifts than the frequency resolution of the semi-coherent methods used here.~\cite{Mukherjee:2017qme}. Accordingly, searches with observing duration of months to less than a few years may safely neglect these effects

Furthermore, as argued in \cite{Covas:2019jqa}, for observing durations of not more than a few years, CW searches targeting NS in binary systems do not require explicit searches of spindown parameters. Consequently, we do not search over spin frequency derivatives in this analysis. Nonetheless, the search retains sensitivity to sources exhibiting spindown or spinup values up to
\begin{equation}
    \left| \dot{f}_0 \right| \le \frac{1}{T_{\text{SFT}} T_{\text{obs}}},
    \label{eq:spindown_limit}
\end{equation}
the value above which the frequency track would deviate by more than one frequency bin in the data during the total observing run, $T_\textrm{obs}$.

\section{\label{sec:data_used} Data Used}

For this search we use data from the first part of the fourth observing run of the LIGO-Virgo-KAGRA detector network. This segment of the run covers the period from 24 May 2023 15:00 UTC (GPS time 1368975618 s) to 16 January 2024 16:00 UTC (GPS time 1389456018 s). Correspondingly, this gives a total observing time of $T_{\rm obs} = 20477806~\mathrm{s}$ or 237~days.

This analysis makes use of data from the LIGO detectors at Hanford (H1) and Livingston (L1)~\cite{LIGOScientific:2014pky}. We do not use data from Virgo~\cite{VIRGO:2014yos}, which was not observing in O4a; nor from KAGRA~\cite{KAGRA:2018plz}, which was not online for most of the run.

During O4a, H1 and L1 were in observing mode 67.5\% and 69.0\% of the time, respectively \cite{Capote:2024rmo}. This represents a lower duty factor than in the O3a \texttt{BinarySkyHough} search \cite{LIGOScientific:2020qhb, 2020PhRvD.102f2003B}. However, the observation time was significantly longer in O4a than in O3a (237~days vs 183~days). Multiplying the observing time by the duty factor, the relative amount of observing data is increased by 20.26\% in O4a compared to O3a.

The data exhibit several different artifacts that can interfere with CW searches, including lines, combs, and hardware injections, any one of which may mimic the behavior of CW signals \cite{2018PhRvD..97h2002C}. Instrumental lines are narrow-band features that appear as persistent monochromatic peaks in the detector's power spectral density. These features can originate from various sources such as injected calibration lines, digital clocks, power supplies, or environmental couplings~\cite{2015CQGra..32c5017E}. These lines can have time dependent, non-stationary behavior in frequency and amplitude, and while some lines are isolated, others form part of broader spectral structures known as \emph{combs}, which consist of multiple harmonics equally spaced in frequency and often arise from a common non-astrophysical origin \cite{2018PhRvD..97h2002C}. 

The search is conducted using SFTs with a baseline duration of $T_\mathrm{SFT} = 1024$ s~\cite{Allen:T040164}. We start from the calibrated detector strain data channel \texttt{GDS-CALIB\_STRAIN\_CLEAN} (C00 calibration, \cite{10.1088/1361-6382/ae1095}), restricted to science-mode segments with CAT1 vetoes~\cite{Goetz:T2400058}. These data are first processed by the self-gating algorithm~\cite{Davis:T2400003}, which removes large transient disturbances in the time domain. The gated strain data (G02) are then processed into Short Fourier Transform Data Base (SFDB) files~\cite{2005CQGra..22S1197A}; this step includes another time-domain cleaning stage to suppress smaller glitches and transient artifacts.

For C00 low-latency data, the O4a uncertainty envelopes (v1) show amplitude and phase systematic errors of approximately $2\%$ and $2^\circ$ across the 20-2000~Hz band~\cite{10.1088/1361-6382/ae1095}. Finally, the cleaned and calibrated SFDBs are converted into SFTs with \texttt{LALSuite}'s \texttt{WriteSFTsfromSFDBs} tool, following the same procedure employed in the O3a all-sky binary CW search~\cite{LIGOScientific:2020qhb, lalsuite}.

Each SFT is produced with a 50\% overlap between consecutive segments and a Tukey window with tapering parameter $\beta_{\mathrm{Tukey}} = 0.5$. Tukey-windowed SFTs contain a flat central region and tapered edges overlapped with consecutive SFTs, retaining high sensitivity to CW-type signals~\cite{2014PhRvD..90d2002A}. This procedure results in 25,426 1024-s SFTs from the H1 data and 26,552 1024-s SFTs from the L1 data.

\section{\label{sec:search_pipeline} Search pipeline}

The search is built using the \texttt{fasttracks} pipeline to enable rapid evaluation of CW detection statistics with GPU acceleration, introduced in~\cite{Tenorio:2024jgc}. In this search, we use a parameter-space partitioning strategy for the six-dimensional search space $\lambda=\left\{ f_{0},\hat{n},a_{p},\Omega,\phi_{\mathrm{b}}\right\} $ (unit vector $\hat{n}$ composed of $\alpha$ and $\delta$) associated with binary CW signals, also introduced in \cite{Tenorio:2024jgc}. The pipeline evaluates detection statistics over several search templates, randomly sampled and in separate parameter-space regions known as \textit{boxes}. This novel parameter-space partitioning naturally leads to a follow-up strategy: in each box of parameter space, only the most significant outliers are followed-up in a more sensitive search, while boxes dominated by instrumental artifacts can be vetoed and excluded from further analysis. Moreover, rather than covering the parameter space with a regular grid of templates, we employ random sampling with resolution chosen to control template mismatch. Details of the resolution criteria will be discussed in Sec.~\ref{sec:partitioning_boxes}, specifically Eq.~\eqref{eq:resolutions}. This approach does not degrade sensitivity compared with grid-based methods, as it keeps the average mismatch between templates and potential signals controlled and reduces correlations between templates~\cite{Allen:2022lqr, Merou:2025ark}.

The output from each box is then passed to a post-processing stage (Sec.~\ref{sec:post_processing}), where high-significance candidates are subsequently followed up using more sensitive methods, as detailed in Sec.~\ref{sec:follow_up}.

\subsection{\label{sec:parameter_space} Parameter space}

We perform a blind, all-sky search for CWs in the frequency range from 100~Hz to 350~Hz. The computational cost of the search grows steeply with frequency, scaling as $f_0^6$ due to the six-dimensional parameter space as derived in~\cite{Tenorio:2024jgc}. As a result, higher-frequency bands are significantly more expensive to analyze than lower ones.

The total frequency range is divided into 0.125~Hz sub-bands, for a total of 2000. Each sub-band is processed independently and partitioned into boxes (Sec.~\ref{sec:partitioning_boxes}).

\begin{figure}[ht]
    \includegraphics[width=0.5\textwidth]{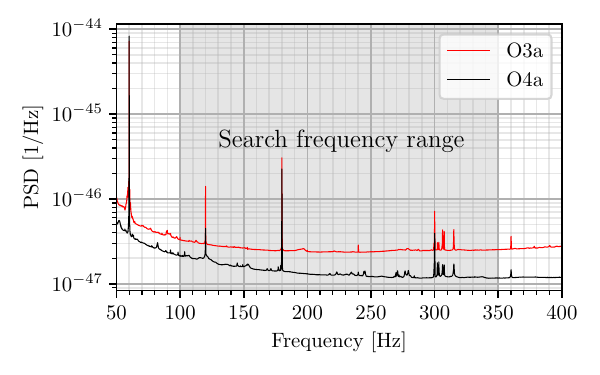}
    \caption{Comparison of the O3a and O4a multi IFO power spectral density using the data from the H1 and L1 LIGO detectors. The grey area indicates the search frequency range.}
    \label{fig:psd_fig} 
\end{figure}

The IFO's sensitivity in each frequency band can be seen in Fig.~\ref{fig:psd_fig}, which illustrates the inverse square root averaged multi-IFO (H1 + L1) power spectral densities (PSDs). We compute the PSD from the single-sided IFO PSD, $S_n(f)$, using an average of inverse-squared PSDs from individual SFTs, as in~\cite{LIGOScientific:2007hnj},
\begin{equation}
S_n(f) = \sqrt{\frac{N}{\sum_{\alpha} \left(S_{\alpha}(f)\right)^{-2}}},
\label{eq:psd}
\end{equation}
where $N$ is the number of SFTs individually indexed by $\alpha$ contributing to the estimate at frequency $f$ and $S_{n}(f)$ is the estimated multi-IFO PSD in the frequency band, computed as the power-2 sum of the individual IFO PSDs. Note that different analyses differ in the mean used to compute the respective PSDs. In the case of CW searches, the power-2 gives the most representative PSD for our searches and for the search sensitivity estimation in Sec.~\ref{sec:sensitivity}.

In O4a, the detectors show improved sensitivity compared to O3a. Whereas in O3a the most sensitive frequencies were in the 150-300~Hz range, in O4a, the optimal performance is in the 200-400~Hz range. The O4a search range overlaps with this most sensitive band of the detectors during the run.

\begin{figure}[ht]
    \includegraphics[width=0.5\textwidth]{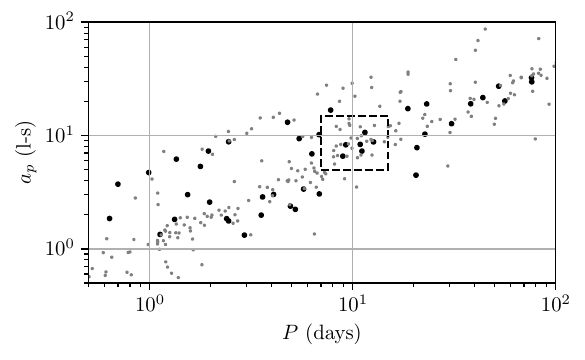}
    \caption{Binary parameter space covered in the search. Large black dots indicate known pulsars with possible CW emission frequencies in the search range, while grey points correspond to sources outside this range. The boxed rectangular region denotes the portion of the binary parameter space explored in this search. Source: ATNF Pulsar Catalogue~\cite{Manchester:2004bp}.}
    \label{fig:binary_param_space}
\end{figure}

For the binary orbital parameters, we target signals from systems with projected semi-major axis amplitude $a_p$ along the line of sight between 5 and 15 light-seconds and orbital periods $P$ ranging from 7 to 15~days. As illustrated in Fig.~\ref{fig:binary_param_space}, this parameter space is chosen to match measured characteristics of observed galactic NSs, focusing on a parameter region which circumscribes a high-density area of known pulsars in binary systems~\cite{Manchester:2004bp}. This choice also corresponds to Region B from the O3a \texttt{BinarySkyHough} search~\cite{LIGOScientific:2020qhb}, and overlaps in the upper 50~Hz with the range in~\cite{Covas:2022rfg} and for $a_p$ in the 10 to 15 l-s range, but not in $P$. The O3a \texttt{BinarySkyHough} search~\cite{LIGOScientific:2020qhb} also included other parameter space regions below 100~Hz; we did not repeat the search in those areas in order to maximize the search sensitivity where the O4a sensitivity had improved the most compared to O3a.

The full search parameter space is summarized in Table~\ref{table:parameter_space}. In addition to the intrinsic frequency and binary orbital parameters discussed above, the search covers the whole sky in $\alpha$ and $\delta$. The binary orbital phase $\phi_{\mathrm{b}}$ is also searched over its entire physical range.

\begin{table}[]
\centering %
\begin{tabular}{ccccccc}
\toprule 
Parameter & $f_{0}$ {[}Hz{]} & $\alpha$ {[}rad{]} & $\delta$ {[}rad{]} & $P$ {[}days{]} & $a_{p}$ {[}l-s{]} & $\phi_{\mathrm{b}}$ {[}rad{]}\tabularnewline
\midrule 
Range & {[}100,350) & $[0,2\pi)$ & $(-\frac{\pi}{2},\frac{\pi}{2})$ & $[7,15)$ & $[5,15]$ & $[0,2\pi)$\tabularnewline
\bottomrule
\end{tabular}\caption{Search parameter ranges.}
\label{table:parameter_space}
\end{table}

\subsection{\label{sec:detection_statistics} Detection statistics}

To identify potential CW signals in the data, we rely on detection statistics that measure the accumulation of excess power along the time–frequency tracks predicted by a given signal model. In practice, a CW signal from a NS in a binary system will produce a characteristic frequency evolution determined by the source parameters. For each template, we define a frequency track across the sequence of SFTs, which represents how a CW signal would evolve in time given the template parameters. Along this track, we add together the corresponding values of the detection statistic from each SFT. If a real signal is present, the detection statistic will accumulate along the track and stand out above the random fluctuations of detector noise, whereas in the absence of a signal the summed statistic will remain consistent with noise.

As in~\cite{Tenorio:2024jgc}, we employ two complementary statistics: the weighted normalized power, which directly measures excess power in the data, and the Hough weighted number count, which is more useful to evaluate the persistence of such excess power across the observing time. The derivation below introduces the normalized power statistic and shows how it is summed along signal tracks to construct the detection statistic used in the search.

Following the notation convention in~\cite{Tenorio:2024jgc}, let $\tilde{x}\left[t;f\right]$ denote the value of the SFT at time $t$ (the start time of the SFT) and frequency $f$, each SFT is identified by an $i$ index and the detector index $X$. Normalizing by the single-sided noise PSD in the band under analysis $S_{nXi}\left[f\right]$ (The PSD is estimated with a 101-bin running-median estimate for each SFT and frequency $f$) and the SFT time baseline $\Tsft$, we define the normalized power statistic as
\begin{equation}
    s\left[t;f\right]=\frac{4\left|\tilde{x}\left[t;f\right]\right|^{2}}{\Tsft S_{nXi}\left[f\right]}.
    \label{eq:normalized_power}
\end{equation}
A search template $\lambda=\left\{ f_{0},\hat{n},a_{p},\Omega,\phi_{\mathrm{b}}\right\}$ defines a signal frequency evolution  $f\left(t_{};\lambda\right)$ across the observing time. The summed weighted normalized power detection statistic is computed by summing the normalized power along the signal track defined by the template,
\begin{equation}
    s(\lambda) = \sum_{X,i} w_{Xi}(\hat{n}) s[t_{Xi}; f(t_{Xi}; \lambda)] \,,
\end{equation}
where $w_{Xi}(\hat{n})$ are time-dependent weights that incorporate both detector sensitivity and antenna response, $t_{Xi}$ refers to time at index $i$ and detector $X$. The appropriate weights for all-sky searches from isotropically oriented sources are \cite{Palomba:2005fp, LIGOScientific:2007hnj}
\begin{equation}
    w_{Xi}(\hat{n}) \propto \frac{a^2(t_{Xi}; \hat{n}) + b^2(t_{Xi}; \hat{n})}{S_{nXi}},
    \label{eq:weights}
\end{equation}
where $S_{nXi}$ is the averaged single-sided PSD of the noise in the frequency band under analysis\footnote{Note the difference between $S_n(f)$ in Eq.~\eqref{eq:psd} (multi-IFO PSD computed as using average of inverse-squared PSD), the $S_{nXi}\left[f\right]$ in Eq.~\eqref{eq:normalized_power} (estimated with a 101-bin running-median estimate for each SFT and frequency $f$) and the $S_{nXi}$ in Eq.~\eqref{eq:weights} (averaged single-sided PSD of the noise in the frequency band under analysis).}, and where $a$ and $b$ are the detector's response functions to the two polarizations of the GW for a source located in direction $\hat{n}$ on the sky. Their definition can be found at~\cite{Jaranowski:1998qm}. These weights are normalized such that
\begin{equation}
\sum_{X,i}w_{Xi}\left(\hat{n}\right)=1.
\label{eq:weights_normalization}
\end{equation}

In order to evaluate the detection statistics across large template banks, we use the vectorization of the computation over batches of templates as implemented in \texttt{fasttracks}~\cite{Tenorio:2024jgc,fasttracks}. This vectorization is done using the \texttt{jax.vmap} function~\cite{jax2018github}.

In addition to the normalized power statistic, we also use the weighted Hough number count~\cite{Krishnan:2004sv} to filter out transient noise artifacts. This statistic was previously also used in the O3a \texttt{BinarySkyHough} search~\cite{LIGOScientific:2020qhb} and has been used in a multitude of CW searches~\cite{2014PhRvD..90d2002A,Covas:2020nwy,LIGOScientific:2019yhl, KAGRA:2022dwb}. It is defined as
\begin{equation}
\text{nc}\left(\lambda\right)=\sum_{X,i} w_{Xi}(\hat{n})\text{nc}\left[t_{Xi}; f\left(t_{Xi};\lambda\right)\right],
\end{equation}
where the number count of an SFT is defined as the digitized normalized power per SFT given by
\begin{equation}
\text{nc}\left[t,f\right]=\begin{cases}
1 & \text{if }s\left[t;f\right]>3.2\\
0 & \text{otherwise}
\end{cases}.
\end{equation}
As in~\cite{Tenorio:2024jgc}, the threshold $s\left[t;f\right]>3.2$ is equivalent to the $\rho_{\text{th}}>1.6$ optimal threshold to minimize the false dismissal rate derived in \cite{Krishnan:2004sv}, given our definition of normalized power in Eq.~\eqref{eq:normalized_power}.

The Hough number count acts as a persistence filter. While transient noise may produce large spikes in the normalized power statistic $s\left[t;f\right]$, it may affect only a small subset of the time series. In contrast, a CW signal will maintain consistent power increases over many segments. Hence, the Hough number count is more robust to brief transients and provides a discriminator against them.

\subsection{\label{sec:partitioning_boxes} Parameter space partitioning into boxes}

To explore the six-dimensional parameter space $\lambda=\left\{ f_{0},\hat{n},a_{p},\Omega,\phi_{\mathrm{b}}\right\} $ of CW signals from NSs in binary systems, we adopt a box-based partitioning strategy, first introduced in~\cite{Tenorio:2024jgc}. Partitioning the parameter space makes post-processing more straightforward, since contaminated boxes can be vetoed and promising candidates followed up independently. Unlike previous searches, which required clustering strategies to sieve through the vast number of initial candidates and group together those likely originating from the same source, our approach reduces the results to a smaller set without this need for clustering.

The idea is to divide the parameter space into a large number of non-overlapping  regions (or ``boxes"), each covering a small portion of the total search volume. Each box is searched independently using a random template bank that densely samples its region.

To determine the size and number of templates in each box, we use characteristic resolution equations. For a box centered at parameters $(f_0, \hat{n}, a_p, \Omega, \phi_{\mathrm{b}})$, the corresponding step sizes in each parameter direction are 
\begin{equation}
    \begin{split}\delta f_{0} & =\frac{1}{T_{\text{SFT}}}\;,\quad\delta\text{sky}=\left(\frac{1}{T_{\text{SFT}}f_{0}v/c}\right)^{2}\;,\quad\delta a_{p}=\frac{1}{T_{\text{SFT}}f_{0}\Omega}\;,\quad\\
    & \delta\Omega=\frac{1}{T_{\text{SFT}}f_{0}a_{p}\Omega T_{\text{obs}}}\;,\quad\delta\phi_{\mathrm{b}}=\frac{1}{T_{\text{SFT}}f_{0}a_{p}\Omega}\:.
    \end{split}
    \label{eq:resolutions}
\end{equation}

To adjust the number of templates per box, we apply an oversampling factor, increasing or decreasing the step sizes defined in Eq.~\eqref{eq:resolutions} times the oversampling factor. Increasing this factor makes the template bank denser (and search potentially more sensitive), however, also more computationally expensive. The oversampling values are tuned based on frequency band, as described in Sec.~\ref{sec:search_steps}. 

The box-based search proceeds as follows:

\begin{enumerate}
    \item Divide the search frequency range into 0.125~Hz sub-bands.
    \item The boundaries of each search box in the physical parameter space $\Delta\lambda$ are mapped to a properly defined auxiliary parameter space, in which the grid steps defined above become uniform (see App.~\ref{sec:partitioning} for the definition of the auxiliary parameter space and the construction of the uniform grid there).
    \item In this uniform space, divide the search volume into equally sized, non-overlapping boxes.
    \item Randomly sample the templates from a uniform distribution inside each box. The number of templates inside a box is the oversampling factor times the amount of templates we would get with a grid if the step sizes were those in Eq.~\eqref{eq:resolutions}.
    \item Map each sampled template back to physical parameter space, evaluate its detection statistics, and retain the template with the highest value according to the criteria in Sec.~\ref{sec:search_steps}.
\end{enumerate}

\begin{figure}[ht]
    \includegraphics[width=0.5\textwidth]{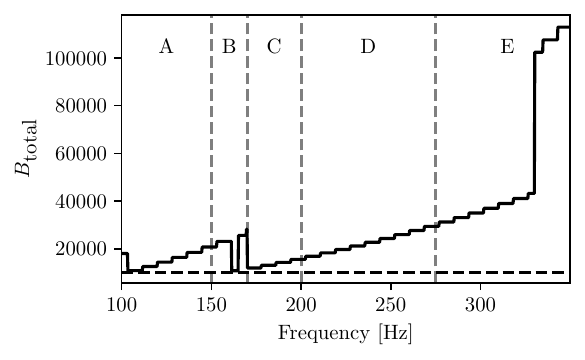}
    \caption{
    Total number of boxes per 0.125~Hz frequency band across the full search range. The lettered regions (A-E) correspond to the frequency intervals defined in Table~\ref{table:frequency_regions}. The dashed horizontal line marks the minimum enforced number of boxes ($10^4$). Sudden changes in the number of boxes correspond to transitions between regions where either the minimum box constraint or the $10^8$ templates-per-box criterion dominates the partitioning as explained in App.~\ref{sec:partitioning}.}
    \label{fig:total_boxes}
\end{figure}

The number of boxes used per frequency band is shown in Fig.~\ref{fig:total_boxes}, and further details of the partitioning algorithm can be found in App.~\ref{sec:partitioning}.

\subsection{\label{sec:search_steps} Two step setup}

Our search adopts a two-step strategy inspired by the methodology used in the O3a \texttt{BinarySkyHough} search~\cite{LIGOScientific:2020qhb}.

For each box in the parameter space, we begin by evaluating the weighted Hough number count statistic $\text{nc}(\lambda)$ for all templates using only half of the available SFTs, skipping every other SFT in the time series. This first step acts as a filter to identify templates associated with CW signal behavior, showing persistent power accumulation over time. We use half the SFTs in order to reduce the computational cost of this first step.

From this first-stage evaluation, we retain the top 0.2\% templates per box ranked by $\text{nc}(\lambda)$. This fraction corresponds approximately to those templates whose $\text{nc}(\lambda)$ exceeds the mean by more than three standard deviations, assuming Gaussian noise. The mean and standard deviation are calculated from all the $\text{nc}(\lambda)$ computed per box.

In the second step of the search, we compute the weighted normalized power statistic $s(\lambda)$ for the subset of templates retained from the first step using the full set of SFTs. For each box, we identify the template with the highest $s(\lambda)$ as the most significant candidate from that region of parameter space.

Different boxes contain different weight distributions $w_{Xi}(\lambda)$ because they can be located at different sky positions. Therefore, to allow for sky-independent statistical significance comparisons across boxes, we use the weighted normalized power critical ratio (CR), defined as
\begin{equation}
\Psi_s = \frac{s - \langle s \rangle}{\sigma_s},
\end{equation}
where $\langle s \rangle$ and $\sigma_s$ are the expected mean and standard deviation of the weighted normalized power statistic in Gaussian noise. These are computed analytically for each box from the weights following the derivation in~\cite{Tenorio:2024jgc},
\begin{align}
    \langle s \rangle &= 2 \sum_{X, i} w_{Xi}\,, \\
    \sigma^2_{s} &= 4 \sum_{X, i} w^2_{Xi}\,,
\end{align}
where, given the weights normalization in Eq.~\eqref{eq:weights_normalization}, $\langle s\rangle=2$.

\begin{table}
    \begin{tabular}{cccc}
    \toprule
        Region & Frequency [Hz] & Oversampling & Follow-up CR threshold\\
    \midrule
        A & [100, 150) & 4.0 & 7.0\\
        B & [150, 170) & 2.0 & 7.0\\
        C & [170, 200) & 1.0 & 7.0\\
        D & [200, 275) & 0.25 & 7.0\\
        E & [275, 350) & 0.25 & 7.2\\
    \bottomrule
\end{tabular}
    \caption{Oversampling factor and follow-up CR threshold used in each frequency region.}
    \label{table:frequency_regions}
\end{table}

\begin{figure*}
\includegraphics[width=1\textwidth]{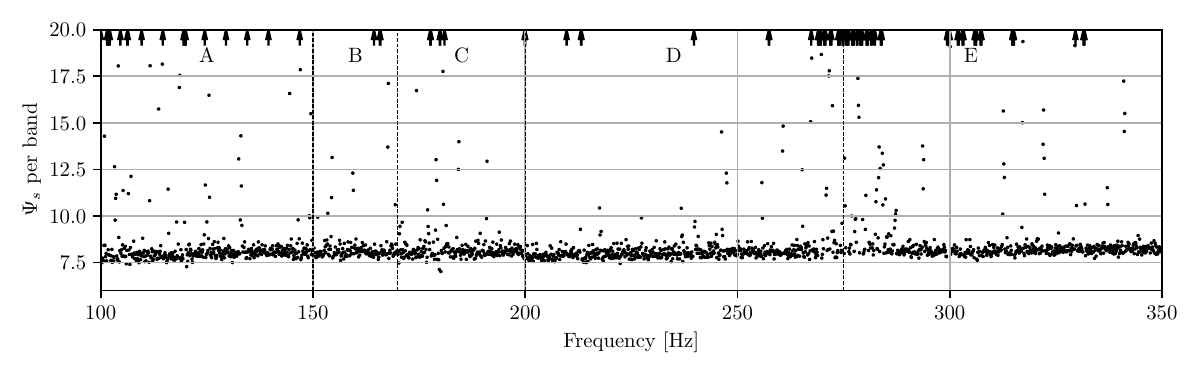}
\caption{Highest CR values per frequency band. Arrows indicate frequency bands where the highest value is above 20.}
\label{fig:crs}
\end{figure*}

We use different oversampling factors across the frequency range. At lower frequencies, where the number of required templates is relatively small, we use higher oversampling values, resulting in denser template banks. Conversely, at higher frequencies, where the parameter space volume increases due to frequency-dependent resolution scaling, we reduce the oversampling factor. 

We define five frequency regions, each with a corresponding oversampling factor and follow-up threshold. Candidates with CR values above the specified threshold are further analyzed in the follow-up stage (see Sec.~\ref{sec:follow_up}). The different frequency regions together with their oversampling factors and follow-up thresholds are described in Table~\ref{table:frequency_regions}. The oversampling factors and follow-up thresholds are chosen based on the available computational resources in order to maximize the search sensitivity to continuous gravitational waves.

Fig.~\ref{fig:crs} illustrates the highest CR values obtained per each frequency band in each region. In some frequency bands, the highest CR values are orders of magnitude larger than those in neighboring bands, a sign of possible contamination from instrumental artifacts. This motivates the introduction of a band-vetoing stage to identify and veto such bands.

\section{\label{sec:post_processing} Known lines veto}

\begin{figure*}[ht]
    \includegraphics[width=\textwidth]{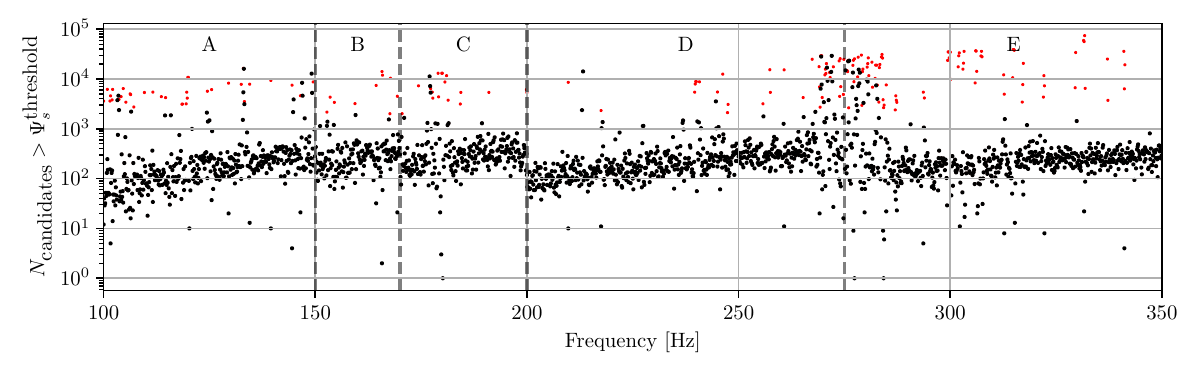}
    \caption{Number of candidates per $0.125$~Hz band with CR above the threshold, as a function of frequency. Frequency regions A through E correspond to those defined in Table~\ref{table:frequency_regions}. Each black dot represents the number of candidates above threshold in a single band. Red dots indicate bands that were fully or partially vetoed due to contamination from known instrumental artifacts listed in the O4a lines list~\cite{Goetz:T2400204}.}
    \label{fig:vetoed_bands}
\end{figure*}

In this step, we identify and discard candidates affected by known instrumental artifacts. This is used to mitigate the contamination from spectral lines that can mimic the expected behavior of CW signals~\cite{Jaume:2024eob}.
 
We use the official list of known instrumental lines from the O4a observing run~\cite{Goetz:T2400204} to identify such contaminations. Each known artifact in the list is examined to determine which frequency ranges affected by its harmonics, based on the tracks searched for each frequency band. The contamination reach of each line is described in App.~\ref{sec:vetoed_bands}.

Based on the contamination, bands are classified as either non-vetoed, partially contaminated, or fully contaminated:
\begin{itemize}
    \item \textbf{Non-vetoed bands}: Bands with fewer than 2000 candidates above the CR threshold are not vetoed. One candidate per box per band is retained for follow-up. Candidates from these bands may still be vetoed due to instrumental artifact contamination after the first stage follow-up, as explained in Sec.~\ref{sec:follow_up}. 
    
    \item \textbf{Fully vetoed bands}: Bands that are entirely within the contaminated region of one or more known lines and that exceed 2000  candidates above threshold are removed from the follow-up. These bands are presumed to be dominated by instrumental artifacts.
    
    \item \textbf{Partially vetoed bands}: Bands that are only partially affected by contamination and that also contain more than 2000 candidate  are cleaned by removing candidates whose boxes fall within the contaminated region. Candidates from the rest of the band are followed-up.
\end{itemize}

The 2000 candidates threshold is an informed choice based on the candidates per band resulting from the search and the distribution observed in Fig.~\ref{fig:vetoed_bands}, where most bands have under 1000 candidates above the corresponding follow-up threshold.

To minimize the false dismissal probability of true astrophysical signals~\cite{Jaume:2024eob}, only frequency regions with  well-documented instrumental disturbances are vetoed outright, and partially contaminated bands are partially vetoed (That is, vetoing only the part of the band that may be contaminated by a known line as defined in in App.~\ref{sec:vetoed_bands}) rather than excluded entirely.

Frequency bands not vetoed by this procedure may still be contaminated by unidentified instrumental artifacts. For example, since the search targets signals whose amplitude is much lower than the detector noise floor, it can be affected by lines that were not seen by the O4a lines analysis.

Across the 2000 search frequency bands, each of width 0.125~Hz, the initial number of candidates above their corresponding CR threshold (Table~\ref{table:frequency_regions}) was 3,253,671. After the known lines veto, the number of surviving candidates is reduced to 927,443, corresponding to 28.5\% of the original sample. Table~\ref{tab:vetoed_bands} summarizes the classification of the bands. Fig.~\ref{fig:vetoed_bands} illustrates the number of candidates above threshold in each band. The vetoed frequency ranges can be found in App.~\ref{sec:vetoed_bands}.

\begin{table}[ht]
\centering %
\begin{tabular}{lccc}
\toprule 
\textbf{Band type\quad} & \textbf{Count\quad} & \textbf{\% Bands\quad} & \textbf{Bandwidth {[}Hz{]}}\tabularnewline
\midrule 
Non-vetoed & 1818 & 90.9\% & 227.25\tabularnewline
Vetoed & 182 & 9.1\% & 16.995\tabularnewline
\quad{}Totally & 65 & 3.25\% & 8.125\tabularnewline
\quad{}Partially & 117 & 5.85\% & 8.87\tabularnewline
\bottomrule
\end{tabular}\caption{Summary of frequency band classification after the application of
line and comb vetoes. Partial vetoes refers to bands where only a
fraction of the templates are discarded due to contamination. The partially vetoed bands sum 14.625~Hz, of which 60.65\% have been vetoed, corresponding to the 8.87~Hz on the table.}
\label{tab:vetoed_bands}
\end{table}

\section{\label{sec:sensitivity} Sensitivity Depth Estimation}

Similar to previous all-sky searches, we assess the sensitivity of this search to putative sources in the search space using software simulated signals added to detector data in selected frequency bands~\cite{LIGOScientific:2019yhl,Covas:2020nwy,LIGOScientific:2017csd,LIGOScientific:2018gpj,LIGOScientific:2020qhb}. Through these simulations, we establish a population-based $h_0$ that is detectable 95\% of the time.

We measure this $h_0$ using the sensitivity depth~\cite{Behnke:2014tma,Dreissigacker:2018afk}, defined as
\begin{equation}
\mathcal{D}=\frac{\sqrt{S_{n}(f)}}{h_{0}},
\end{equation}
where $S_n$ is the single-sided IFO PSD of the detector noise defined in Eq.~\eqref{eq:psd}, and $h_0$ is the intrinsic signal strain amplitude. The sensitivity depth provides a figure of merit that normalizes $h_0$ by the detector noise, facilitating comparisons across different frequency bands and search methods. Nonetheless, for these comparisons it is important to note that the sensitivity depth remains highly dependent on the searched parameter ranges in addition to frequency. These sensitivity depth estimates are used to set thresholds in the follow-up procedure (Sec.~\ref{sec:follow_up}) and to study the astrophysical reach of the search (Sec.~\ref{sec:results}).

For each 0.125~Hz frequency band in which injections are performed, we select the highest $S_n(f)$ within the band to evaluate sensitivity depths and search for the $\mathcal{D}^{95\%}$ value at which we determine the amplitude $h_0$ at which 95\% of injections are recovered above threshold.

For each injected signal, we construct a small region centered on the injection parameters. The region spans $\pm$3 resolution units in all search dimensions according to Eq.~\eqref{eq:resolutions}, except for the binary phase $\phi_{\mathrm{b}}$, where a $\pm$5 resolution range is used. This choice has been made because it is more challenging to constrain the binary phase of simulated signals compared to the other parameters. These regions are designed to be compatible with the first-stage follow-up prior ranges in Sec.~\ref{sec:follow_up}, making sure that any detectable signal in this setup is recovered and followed up by the pipeline.

Besides the resolution intervals shown in Eq.~\eqref{eq:resolutions}, the specific ones for $\alpha$ and $\delta$ are set using 
\begin{align}
\delta\alpha & = \frac{1}{T_{\text{SFT}}(v/c)f_0}\frac{1}{\cos\delta}, \
\delta\delta & = \frac{1}{T_{\text{SFT}}(v/c)f_0},
\label{eq:resolutions_sky}
\end{align}
which take into account the $\cos\delta$ dependence arising from spherical coordinates to make resolution intervals correctly centered and spaced around injected signals.

All signals are injected into the same O4a SFT data used in the search using the \texttt{PyFstat} software package~\cite{PhysRevD.97.103020, Keitel2021, pyfstatzenodo, Wette:2020air}. \texttt{PyFstat} is built on top of \texttt{LALSuite}~\cite{lalsuite} and contains multiple utilities for CW searches, such as injection generation or candidate follow-ups. After injecting the signals, we rerun the whole analysis on the small intervals, computing the detection statistics over the corresponding box with the same oversampling factor as used in the main search. We then check whether the candidate template CR value exceeds the corresponding CR follow-up threshold from Table~\ref{table:frequency_regions}, which allows us to determine if the injection would have been one of the outliers passing to the follow-up stage. Afterwards, we use this information to estimate the detection efficiency and corresponding sensitivity depths.

To classify an injected signal as detected, we require that at least one candidate template within the injection search region satisfies the following three conditions:
\begin{enumerate}
    \item The number count statistic nc$(\lambda)$, computed using half of the SFTs, must be greater than the lowest nc$(\lambda)$ value from the 
    first stage of the search for the box in which the signal was injected. This ensures that the candidate template would have passed the persistence filter used in the first search step.
    \item The CR must exceed the maximum found in the second stage of the search for that same box. This confirms that the candidate template would have been selected as the most significant outlier in that box.
    \item The CR value must also surpass the follow-up CR threshold for the corresponding frequency band.
\end{enumerate}

We inject signals into at least one 0.125~Hz band every 5~Hz across the search frequency range in non-vetoed bands in Sec.~\ref{sec:post_processing}. Therefore, the sensitivity estimates do not apply to the vetoed frequencies from Sec~\ref{sec:post_processing}, listed in Table~\ref{tab:vetoed_freq_ranges}.

In each selected band, we inject 500 signals at different depths centered around the expected $\mathcal{D}^{95\%}$, with parameters drawn from uniform distributions over the phase and amplitude parameters. Each signal is injected at a random frequency inside the band and with random binary orbital parameters within the ranges defined in Table~\ref{table:parameter_space}. In order to estimate the sensitivity depth at each frequency, we compute the detection efficiency as the proportion of detectable injections in each depth, and estimate the sensitivity using sigmoidal fits and interpolating at which sensitivity depth 95\% of the injections are tagged as detected. The procedure is detailed in App.~\ref{sec:app_sensitivity}.

\begin{figure}
\includegraphics[width=0.5\textwidth]{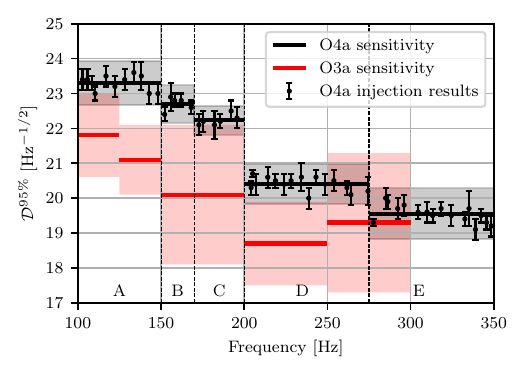}
\caption{$95\%$ sensitivity depths $\mathcal{D}^{95\%}$ as a function of frequency. Each point represents a 0.125~Hz band with injection results, with error bars showing $\pm\sigma$ from the sigmoidal fit. Shaded regions show the average sensitivity depth per frequency region with its 3$\sigma$ uncertainty (see Table~\ref{table:average_depths}). Comparison between the results of this search (in black) and the results of the O3a BinarySkyHough search~\cite{LIGOScientific:2020qhb} (in red).}
\label{fig:depth_plots}
\end{figure}

Fig.~\ref{fig:depth_plots} shows the results for sensitivity depth estimates as a function of frequency for the different selected frequency bands. The mean value across each frequency region (as defined in Table~\ref{table:frequency_regions}) is computed from these values and is summarized in Table~\ref{table:average_depths}.

\begin{table}[ht]
\centering %
\begin{tabular}{ccc}
\toprule 
Region  & Frequency {[}Hz{]} & $\langle\mathcal{D}^{95\%}\rangle\pm3\sigma\;[\text{Hz}^{-1/2}]$\tabularnewline
\midrule 
A  & {[}100, 150) & $23.3\pm0.6$ \tabularnewline
B  & {[}150, 170) & $22.7\pm0.5$ \tabularnewline
C  & {[}170, 200) & $22.2\pm0.4$ \tabularnewline
D  & {[}200, 275) & $20.4\pm0.6$ \tabularnewline
E  & {[}275, 350) & $19.6\pm0.7$\tabularnewline
\bottomrule
\end{tabular}\caption{Average $95\%$ sensitivity depths across the frequency regions defined in Table~\ref{table:frequency_regions}. The uncertainties correspond to $\pm3\sigma$ from the sigmoidal fit.}
\label{table:average_depths} 
\end{table}

\section{\label{sec:follow_up} Search outliers follow-up}

Candidates with CRs above the follow-up thresholds (Table~\ref{table:frequency_regions}) in the main stage of the search (see Fig.~\ref{fig:vetoed_bands}) are subjected to a more sensitive follow-up analysis. The follow-up analysis is performed using the \texttt{PyFstat} software package~\cite{PhysRevD.97.103020, Keitel2021, pyfstatzenodo, Wette:2020air}, as was done in the O3a \texttt{BinarySkyHough} search~\cite{LIGOScientific:2020qhb}. \texttt{PyFstat} utilizes the parallel-tempered Markov Chain Monte Carlo (PTMCMC) sampler provided by the \texttt{ptemcee} package for the candidate follow-up~\cite{2013PASP..125..306F, 2016MNRAS.455.1919V}. 

This stage uses longer coherence times and includes additional frequency-evolution parameters. Increasing coherence times may also reject potential signals that do not conform to the phase model enforced by the longer coherence lengths~\cite{Carlin:2025sxm}. Signals whose spindown is larger than Eq.~\eqref{eq:spindown_limit} or whose eccentricity is larger than Eq.~\eqref{eq:eccentricty} are expected to already be lost during the search stage and their high spindowns or eccentricities are not considered inside the parameter ranges of this follow-up stage. Changes to the frequency evolution model due to spin-wandering are not considered in the follow-up.

Increasing $T_{\rm coh}$ strengthens real CWs relative to Gaussian noise because correctly demodulated signals accumulate coherent power with time, while instrumental artifacts, which lack the astrophysical Doppler modulations, are typically diminished by the demodulation and are further rejected by our prominence-crossing and line-list vetoes. 

Candidates are analyzed using $\mathcal{F}$-statistic searches \cite{Jaranowski:1998qm, Prix:2011qv}. The $\mathcal{F}$-statistic compares data to signal templates in the frequency domain using coherent integration. The semi-coherent version, the segment-wise $2\hat{\mathcal{F}}$, is constructed by summing the coherent $2\tilde{\mathcal{F}}$ statistic over $N_\textrm{seg}$ data segments, each with duration $T_\text{coh}$,
\begin{equation}
2\hat{\mathcal{F}}\left(\lambda\right)=\sum_{i=0}^{N_{\text{seg}}-1}2\tilde{\mathcal{F}}_{i}\left(\lambda\right).
\end{equation}

In the follow-up process, the MCMC algorithm takes samples using 300 walkers each one with 300 steps to identify and characterize potential maxima, which are then used to  determine whether a candidate passes a detection threshold. The MCMC-based follow-up and its application to CW searches is detailed in~\cite{Ashton:2018ure}.

In the first stage of the MCMC follow-up, candidates from the initial search are analyzed using a significantly increased coherence time to improve sensitivity. Specifically, we increase the coherence time from the $T_{\rm SFT} = 1024$ s used in the main search to $T_{\rm coh} = 12$ hours for the follow-up. This configuration yields a total of $N_\textrm{seg} = 474$ segments over the full observing time. The follow-up priors and MCMC sampler configuration is developed in App.~\ref{sec:followupappendix}.

\begin{figure*}[ht]
\centering
\includegraphics[width=\textwidth]{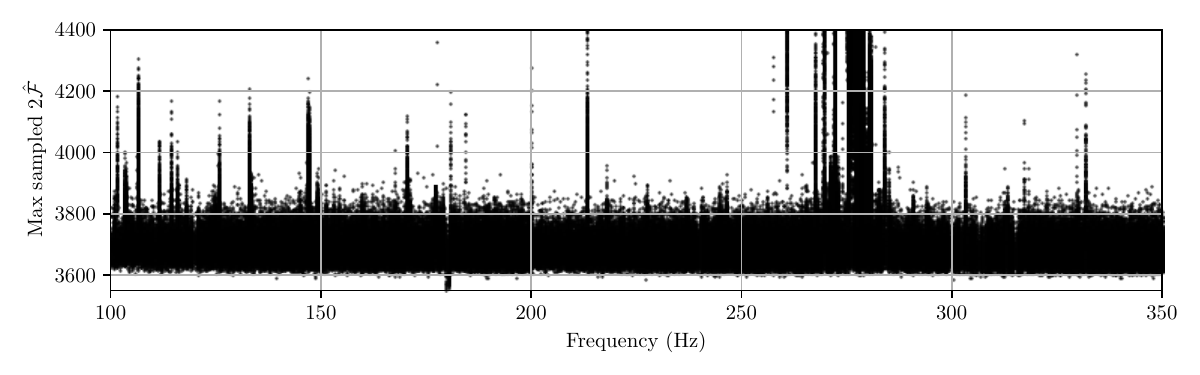}
\caption{Results of the first-stage MCMC follow-up. The figure shows the maximum $2\hat{\mathcal{F}}$ values recovered for each candidate, plotted as 
a function of frequency.}
\label{fig:1st_stage_results}
\end{figure*}

Fig.~\ref{fig:1st_stage_results} summarizes the results of the first-stage MCMC follow-up across all frequency bands, showing the maximum recovered $2\hat{\mathcal{F}}$ values as a function of frequency. As seen, most frequency bands show minimums around $2\hat{\mathcal{F}}=3600$ and maximums around $2\hat{\mathcal{F}}=3900$. However, there are bands where the maximum sampled $2\hat{\mathcal{F}}$ is much higher above 3900. A large number of them appear near the instrumental disturbances around 270~Hz, which indicates that instrumental lines can highly increase the sampled $2\hat{\mathcal{F}}$.

To determine which candidates should advance to a more sensitive second follow-up stage, we calibrated a threshold on the cumulative $2\hat{\mathcal{F}}$ using simulated signal injections. These injections were generated with the same setup and randomized amplitudes described in the Sec~\ref{sec:sensitivity}. The simulated signals were injected at the sensitivity depths values obtained per each region shown in Table~\ref{table:average_depths}. For each injected signal, we recorded the maximum cumulative $2\hat{\mathcal{F}}$ value across all walkers within the corresponding frequency band. By repeating this process for many injections across different frequency bands, sampling at least one band every 5~Hz and covering all regions listed in Table~\ref{table:frequency_regions} at the depths given in Table~\ref{table:average_depths}, we built up the distribution of recovered $2\hat{\mathcal{F}}$ values, which was then used to set the threshold. After running 14750 injections, the detection threshold was then set at $2\hat{\mathcal{F}} = 3850$, chosen such that fewer than 0.013\% of the injected signals (2 of the 14750) would fall below it.

Only injections that were tagged as detected by the search pipeline were followed-up to determine this threshold. Furthermore, the injection follow-up was performed with the same MCMC sampler configuration and priors for the search candidates.

\begin{figure}[ht]
\centering
\includegraphics[width=0.5\textwidth]{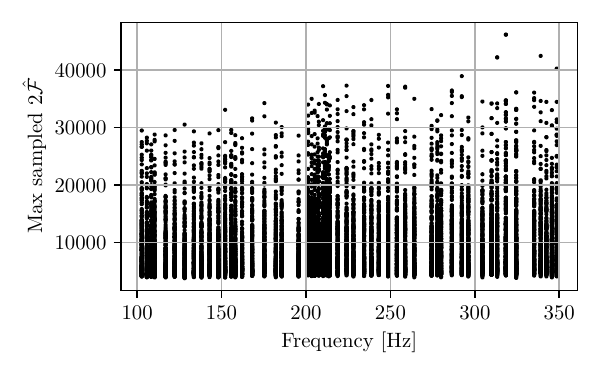}
\caption{First-stage MCMC follow-up results on injection studies with $T_\textrm{coh}=12$ h. Each point corresponds to the maximum $2\hat{\mathcal{F}}$ value sampled for each injected signal. 14750 Injections were run, with 2 obtaining a final $2\hat{\mathcal{F}}$ below 3850. Signals were injected at the sensitivity depths values obtained per each region shown in Table~\ref{table:average_depths}.}
\label{fig:1st_stage_injections}
\end{figure}

The injection results are shown in Fig.~\ref{fig:1st_stage_injections}. Candidates passing this initial $2\hat{\mathcal{F}}$ cut are then subjected to a second follow-up stage with a longer coherence time,  following the methodology outlined in~\cite{PhysRevD.97.103020, Tenorio:2021njf, PhysRevD.110.124049}.

After running the first-stage MCMC follow-up and applying the cumulative $2\hat{\mathcal{F}}$ threshold, several outliers remained, as summarized in Table~\ref{table:followup_summary_combined}. These surviving outliers were further analyzed to determine whether they are consistent with astrophysical signals or can be attributed to instrumental origins.

CW signals are expected to accumulate $2\hat{\mathcal{F}}$ linearly over time, adding to the segment-wise $2\tilde{\mathcal{F}}$ in a relatively uniform way across the observation period. However, if a candidate's frequency evolution briefly overlaps with a strong instrumental disturbance or narrow-band spectral line, it may produce a spike in $2\tilde{\mathcal{F}}$ during that specific time window, while remaining at background levels otherwise.

To identify such cases, we examine the segment-wise $2\tilde{\mathcal{F}}$ distributions of surviving candidates. Following the same approach used in the O3a search, we flag any candidates in bands with at least one segment-wise $2\tilde{\mathcal{F}}$ value exceeding 100. These outliers are classified as  ``prominence-crossing" cases and are likely caused by transient overlaps with lines or short-lived artifacts rather than persistent astrophysical signals. All candidates within-bands that are tagged as prominence-crossing are discarded in this stage and not followed-up.

Notice that this prominence-crossing veto is necessary even if the number-count statistic has previously been used in the candidate selection. Using the number-count during the search provides more importance to candidates consistent with high power across the observing run. Nonetheless, it is not a filter that eliminates all candidates whose high power is a consequence of crossing paths with an instrumental artifact.

\begin{figure}[ht]
\centering
\includegraphics[width=0.5\textwidth]{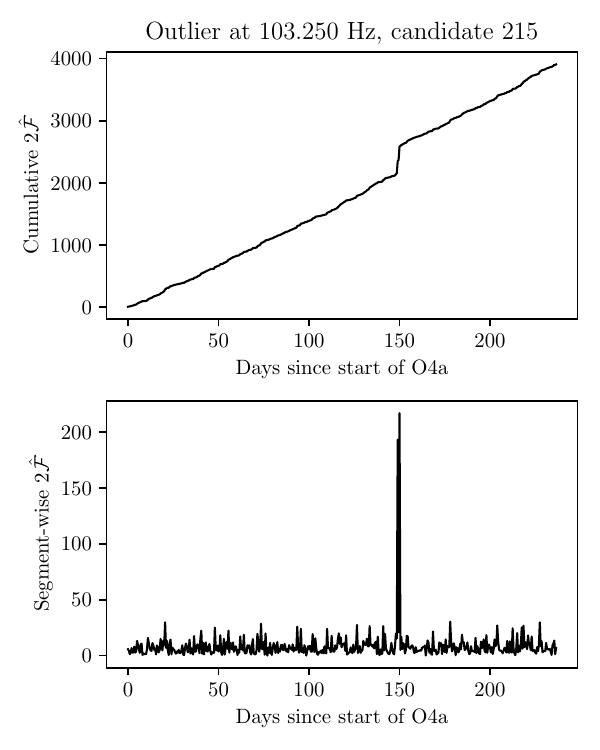}
\caption{Example of a prominence-crossing outlier. The candidate shows a large spike in $2\tilde{\mathcal{F}}$ near day 150 of the observing run. This spike allows it to exceed the follow-up threshold of $2\hat{\mathcal{F}} = 3850$, but the narrow change suggests a superposition with an instrumental disturbance. This candidate is therefore discarded from second-stage follow-up.}
\label{fig:line_crossing}
\end{figure}

An example is shown in Fig.~\ref{fig:line_crossing}, where the candidate band exhibits a sharp and isolated $2\tilde{\mathcal{F}}$ spike near day 150 of the run. While the total $2\hat{\mathcal{F}}$ surpasses the threshold, the strong deviation in a single segment suggests a transient artifact rather than a genuine signal, and the candidate is therefore discarded.

\begin{figure}[ht]
\centering
\includegraphics[width=0.5\textwidth]{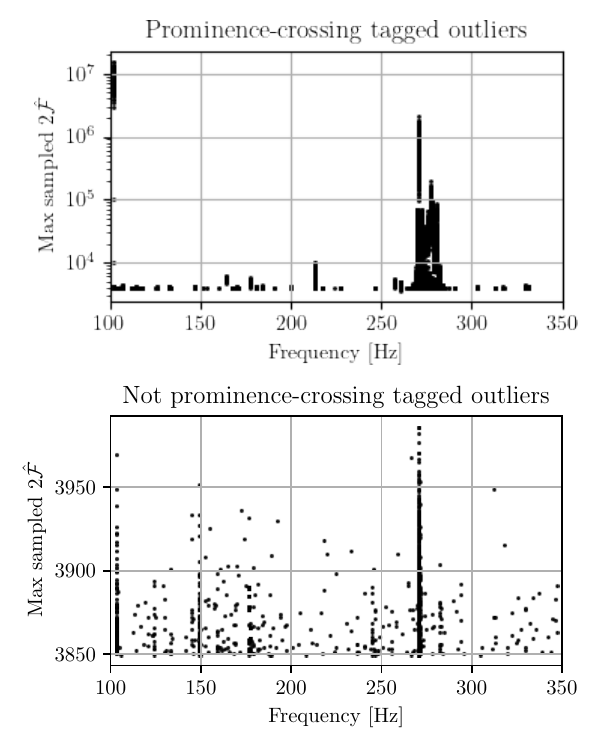}
\caption{Distribution of maximum $2\hat{\mathcal{F}}$ values for surviving outliers. The top panel shows outliers tagged as prominence-crossing and discarded due to transient spectral contamination. These candidates tend to cluster in a few frequency bands and exhibit extremely high $2\hat{\mathcal{F}}$ values. The bottom panel shows the remaining outliers, which were not flagged as line-crossing and are retained for further analysis.}
\label{fig:line_crossing_summary}
\end{figure}

The broader distribution of such outliers is summarized in Fig.~\ref{fig:line_crossing_summary}. The top panel shows those candidates that have been tagged as prominence-crossing and rejected; they tend to be clustered within a few narrow frequency ranges and exhibit significantly elevated $2\hat{\mathcal{F}}$ values. The bottom panel shows the remaining 1,997 outliers, which passed all veto criteria and are retained for further second-stage follow-up analysis. The results from this prominence crossing veto are summarized in Table~\ref{table:followup_summary_combined}.

\begin{table}[ht]
\centering
\begin{tabular}{lc}
\toprule
\textbf{Follow-up Summary} & \textbf{Count} \\
\midrule
Total candidates followed up & 927,443 \\
Candidates surviving $2\hat{\mathcal{F}} > 3850$ & 251,717 \\
\quad - Prominence crossings (discarded) & 249,720 \\
\quad - Surviving to 2nd stage & 1,997 \\
\bottomrule
\end{tabular}
\caption{Summary of candidate outcomes from the first-stage MCMC follow-up. Of the 927,443 candidates analyzed, 251,717 exceeded the threshold $2\hat{\mathcal{F}} > 3850$, yielding a survival rate of 27.14\%. Among these, 249,720 were tagged as prominence or prominence crossings and discarded, while 1,997 (8\%) were retained for second-stage follow-up.  }
\label{table:followup_summary_combined}
\end{table}

Of the 251,717 surviving candidates in Table~\ref{table:followup_summary_combined}, 84\% come from the frequency range between 260~Hz and 280~Hz, a range with a high amount of high-CR results as seen in Fig.~\ref{fig:crs} where there is also a high amount of vetoed bands as seen in Fig.~\ref{fig:vetoed_bands}. It appears that there is a high amount of broad artifacts in the range, not all of them vetoed in Sec.~\ref{sec:post_processing}.

The second-stage follows the multi-stage procedure established in previous works~\cite{PhysRevD.97.103020, Tenorio:2021njf, PhysRevD.110.124049}. In this second follow-up stage, we use a longer coherence time of $T_\textrm{coh}=26$ h, which increases the sensitivity of the search and improves the accuracy of parameter estimation. The rationale for this choice, together with the details of the priors and analysis setup, which runs again 300 walkers with 300 steps each, is provided in App.~\ref{sec:followupappendix}.

\begin{figure}[ht]
\includegraphics[width=0.5\textwidth]{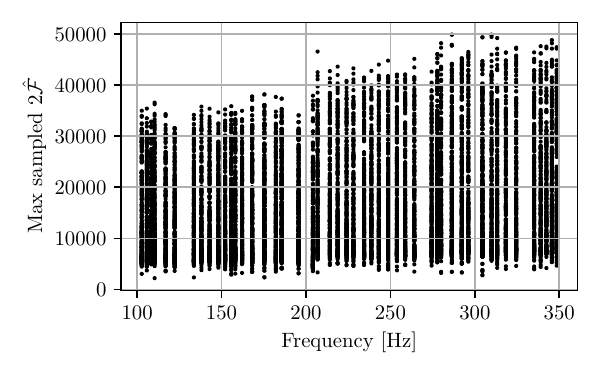}
\caption{Second-stage MCMC follow-up results on injection studies $T_\textrm{coh}=26$ h. Each dot represents the maximum sampled $2\hat{\mathcal{F}}$ value for a detected injection across multiple frequency bands. A total of 14000 injections are shown. Signals were injected at the sensitivity depths values obtained per each region shown in Table~\ref{table:average_depths}.}
\label{fig:2nd_stage_injections}
\end{figure}

The second stage threshold for further follow-ups was calibrated using software injections as in the first stage. Fig.~\ref{fig:2nd_stage_injections} shows the maximum $2\hat{\mathcal{F}}$ values sampled by the MCMC for each injected signal. The lowest sampled value for detectable signals was close to 2200, indicating that a threshold of 2100 would give a false dismissal rate below 1/14000, given the 14000 injections run.

\begin{figure}[ht]
\includegraphics[width=0.5\textwidth]{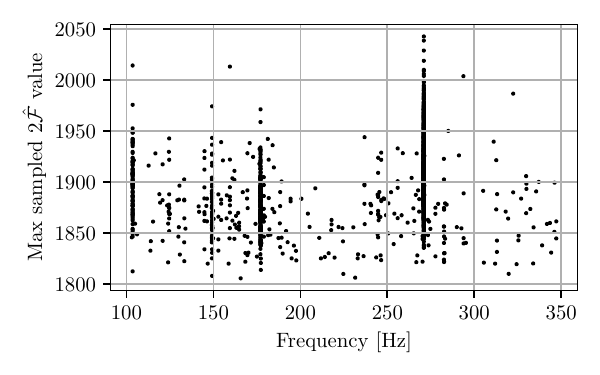}
\caption{Maximum $2\hat{\mathcal{F}}$ values recovered from the second-stage MCMC follow-up for each of the 1,997 candidates. No outlier exceeds the detection 
threshold.}
\label{fig:2nd_stage_results}
\end{figure}

The second-stage pipeline was applied to the 1,997 candidates that survived the first stage. The results are shown in Fig.~\ref{fig:2nd_stage_results}. None of the recovered maximum $2\hat{\mathcal{F}}$ values exceeded the calibrated threshold of 2100. As seen again in Fig.~\ref{fig:vetoed_bands}, there is a high amount of instrumental disturbances around 270~Hz which have been vetoed in this stage. All recovered values were well below the injection  minimum of 2200, and thus none of these outliers are consistent with a detectable CW signal.

Since no candidate survives the second-stage threshold, no further follow-up stages have been carried out. The follow-up phase of the search concludes with no candidate consistent with an astrophysical CW source.

\section{\label{sec:results} Discussion}

This analysis is the first large-scale application of the \texttt{fasttracks} pipeline~\cite{Tenorio:2024jgc, fasttracks}, designed to compute detection statistics with GPU acceleration using SFT-type data. To address the large parameter space of the search, we used a novel box-based partitioning strategy also introduced in~\cite{Tenorio:2024jgc}, in which each parameter-space partition, or box, is analyzed independently. This novel partitioning makes post-processing more straightforward, since contaminated boxes can be vetoed and candidates from each box followed up independently. Whereas previous searches required clustering strategies to filter large numbers of candidates and group those likely originating from the same source, our approach achieves comparable reduction in a simpler manner.

A two-step search approach has been used, with a Hough number count filter in the first step to prioritize candidates showing persistent power accumulation. Using the O4a lines list, frequency bands were checked for line contamination and those with more than 2000 candidates and known line contamination were vetoed.

The surviving 27.14\% of candidates after the known lines veto were analyzed using an MCMC follow-up method using the \texttt{PyFstat} package~\cite{PhysRevD.97.103020, Keitel2021, pyfstatzenodo, Wette:2020air} and $T_\textrm{coh}=12$~h. These candidates were checked to determine if their high statistical values were due to prominence-crossing behavior. Those deemed as not prominence-crossing (8\%) were re-analyzed with a second stage follow-up, which concluded the search with no candidates consistent with astrophysical CW sources. 

\begin{figure*}[ht]
\includegraphics[width=\textwidth]{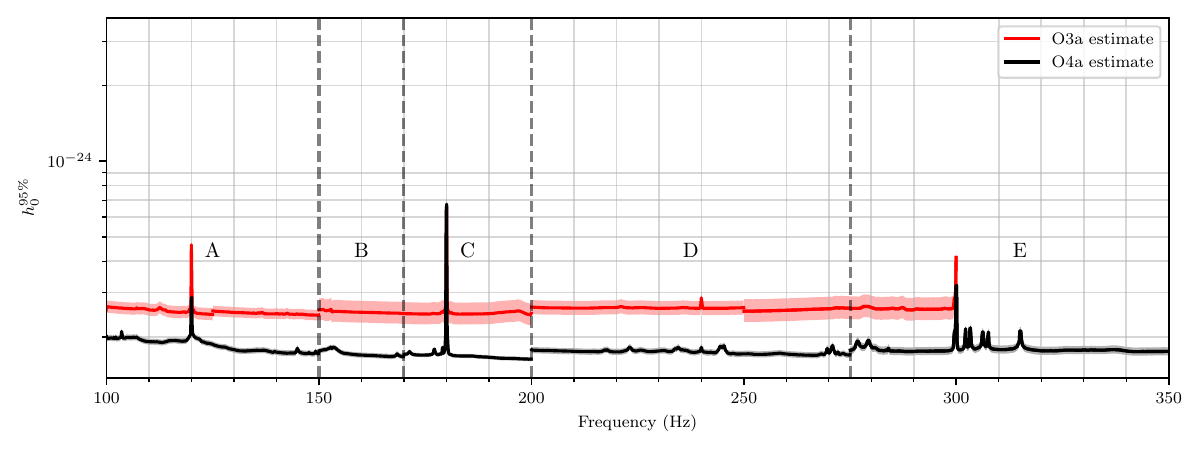}
\caption{Estimated $95\%$ strain amplitude sensitivity $h_0^{95\%}$ derived from the sensitivity depths in Table~\ref{table:average_depths}, using the inverse square root of the averaged PSD. Comparison between the O3a BinarySkyHough search results~\cite{LIGOScientific:2020qhb} (in red), and the results of this present search (in black).}
\label{fig:amplitudes}
\end{figure*}

To evaluate the sensitivity of our search, we conducted injection campaigns across all frequency regions, as described in Sec.~\ref{sec:sensitivity}. From these studies, we derived the $95\%$ confidence strain amplitude sensitivity $h_0^{95\%}$, computed from the estimated sensitivity depths and the noise power spectral density shown in Fig.~\ref{fig:psd_fig}. The resulting sensitivity curve is presented in Fig.~\ref{fig:amplitudes}. The best performance was achieved at $199.9$~Hz, with $h_0^{95\%} = (1.62 \pm 0.3) \times 10^{-25}$, corresponding to an improvement by a factor 1.5 compared to the lowest amplitude obtained by the previous \texttt{BinarySkyHough} search on O3a data~\cite{LIGOScientific:2020qhb}. This improvement over the O3a results is consistent with the expectations given the improvement in search sensitivity shown in Fig.~\ref{fig:psd_fig} and longer observing times plus the improvement in sensitivity depth shown in Fig.~\ref{fig:depth_plots}. These estimates are not applicable to the vetoed frequency ranges listed in Table~\ref{tab:vetoed_freq_ranges}. As in~\cite{Covas:2019jqa}, the reported sensitivities are valid for eccentricities below the maximum value defined in Eq.~\eqref{eq:eccentricty}, above which the signal track would deviate by more than one frequency bin.

\begin{figure}[ht]
\centering
\includegraphics[width=0.48\textwidth]{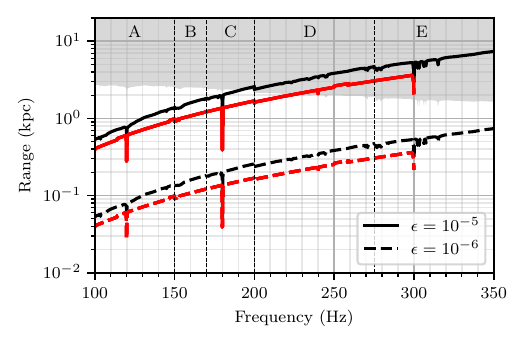}
\caption{Estimated distance reach for sources emitting at $h_0^{95\%}$, based on the sensitivity of the search. Shaded regions indicate distances excluded by the limit imposed by the maximum spindown used in the analysis. Comparison between the O3a BinarySkyHough search results~\cite{LIGOScientific:2020qhb} (in red), and the results of this present search (in black).}
\label{fig:astroreach_distance}
\end{figure}

\begin{figure}[ht]
\centering
\includegraphics[width=0.48\textwidth]{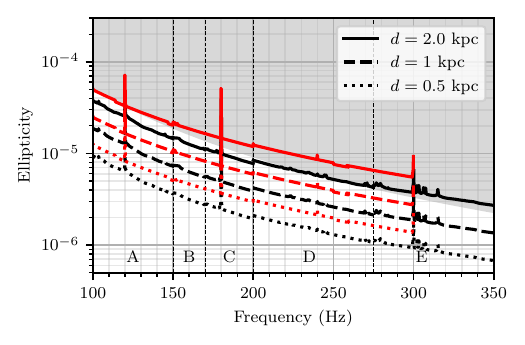}
\caption{Corresponding equatorial ellipticities estimated from the $h_0^{95\%}$ sensitivity, assuming a canonical NS moment of inertia $I_{zz} = 10^{38}$ kg$\cdot$m$^2$. The shaded regions are excluded by the maximum spindown constraint. Comparison between the O3a BinarySkyHough search results~\cite{LIGOScientific:2020qhb} (in red), and the results of this present search (in black).}
\label{fig:astroreach_ellipticity}
\end{figure}
 
Fig.~\ref{fig:astroreach_distance} illustrates the astrophysical reach of this search (in black) derived from the estimated $h_0^{95\%}$ sensitivities and its comparison with the O3a BinarySkyHough search results~\cite{LIGOScientific:2020qhb} (in red). The figure shows the distance range (in kiloparsecs) at which NSs with various ellipticities $\epsilon$ could be detected, assuming the canonical moment of inertia $I_{zz} = 10^{38}$ kg$\cdot$m$^2$. The shaded regions indicate values excluded by being larger than the maximum spindown used in this search (see Eq.~\eqref{eq:spindown_limit}).

Fig.~\ref{fig:astroreach_ellipticity} translates these sensitivity results into constraints on the NS ellipticity $\epsilon$, assuming sources are located at fixed distances. For sources within 1 kpc, the ellipticity can be constrained to below $10^{-5}$ across the majority of the analyzed band (above 130~Hz). These limits begin to probe astrophysically interesting regimes: theoretical expectations for the maximum ellipticity of realistic NSs range from $\sim 10^{-6}$ to $10^{-7}$ for conventional nuclear equations of state, up to $\sim 10^{-5}$ for more extreme or exotic configurations~\cite{Ushomirsky:2000ax, Jones:2024npg}.

Using the ATNF pulsar catalogue~\cite{Manchester:2004bp}, we find that of the 3,473 known pulsars, 715 lie within 2 kpc of Earth, 282 within 1 kpc, and 117 within 0.5 kpc. Of these, 151, 66, and 16, respectively, are in binary systems. This confirms that the sensitivity region of our search overlaps with a non-negligible fraction of the known pulsar population, including binaries. Moreover, restricting the nearby binaries to orbital periods between 7 and 15~days and projected semi-major axes between 5 and 15 light-seconds, we identify 17, 8, and 1 such systems within 2 kpc, 1 kpc, and 0.5 kpc, respectively. These are a relative small number of known pulsars probed by the search. Nonetheless, only a small fraction of the Galactic NS population is observable as radio or gamma-ray pulsars~\cite{Rajwade:2018xxf, Reed:2021scb}, with population synthesis models predicting orders of magnitude more NSs than those currently detected. Consequently, our sensitivity may overlap with a much larger, presently unseen NS population, including systems with properties favorable for the emission of CW. Thus, this search directly probes the parameter space occupied by several known binary pulsars, as well as potentially a much larger population of undetected NSs.

\section{\label{sec:conclusion} Conclusion}

We have conducted an all-sky search for continuous gravitational waves from neutron stars in binary systems using data from the first part of the fourth LIGO-Virgo-KAGRA observing run. The search covered the frequency range from 100 to 350~Hz, targeting systems with projected semi-major axes between 5 and 15 light-seconds and orbital periods between 7 and 15~days.

This work represents the first large-scale application of the \texttt{fasttracks} pipeline. We analyzed O4a data with a two-step procedure: a first Hough number count filter to identify persistent candidates, a second step computing weighted normalized power, followed by an MCMC analysis with extended coherence time follow-ups with \texttt{PyFstat}. Vetoes of line-contaminated bands reduced the candidate set, however, no signals consistent with astrophysical continuous waves were found. Sensitivity estimates from extensive injection campaigns show a minimum detectable amplitude of $h_0^{95\%}=(1.62\pm0.3)\times10^{-25}$ at 199.9~Hz, representing a 1.5-fold improvement over the lowest amplitude achieved by the previous \texttt{BinarySkyHough} search on O3a data~\cite{LIGOScientific:2020qhb}.

The achieved sensitivity covers binary parameters and spin frequencies consistent with those of many known binary pulsars within a few kiloparsecs, as well as with a potentially much larger unseen neutron star population. Constraints on the ellipticity of nearby sources reach below $10^{-5}$, entering the regime predicted by some nuclear equations of state and more exotic neutron star models. The non-detection of continuous gravitational waves at O4 sensitivity imposes strong constraints on neutron stars with exotic compositions within the distance reach of the sources.

Looking ahead, upgrades to the global network of terrestrial gravitational-wave detectors are expected to deliver enhanced strain sensitivity, extended frequency coverage, and longer observing baselines in upcoming observing runs, more so with third-generation detectors such as the Einstein Telescope~\cite{ET:2019dnz} and Cosmic Explorer~\cite{2021arXiv210909882E}. These advancements will increase the chances of detecting CW signals and improve constraints on the galactic neutron star population. Complementary progress will come from next-generation pulsar timing facilities, such as the Square Kilometre Array~\cite{Weltman:2018zrl}, which will expand the known pulsar population and improve the measurement of their evolution, thereby providing valuable priors and targets for gravitational-wave searches.

\begin{acknowledgments}
\input{P2000488_v28.tex}
This paper has been assigned document number LIGO-P2500437.

\end{acknowledgments}

\appendix

\section{\label{sec:partitioning} Parameter-space Partitioning}

To estimate the number of templates required to cover a given region of parameter space, we begin with the resolution steps $\delta\lambda[k]$ given by Eq.~\eqref{eq:resolutions} along each of the parameters in $\lambda=\left\{ f_{0},\hat{n},a_{p},\Omega,\phi_{\mathrm{b}}\right\} $. The total number of templates within a volume $\Delta\lambda$ is given by
\begin{equation}
\mathcal{N}\left(\Delta\lambda\right) = \int_{\Delta\lambda} d\lambda \rho\left(\lambda\right),
\end{equation}
where the local template density, $\rho\left(\lambda\right)$, is defined as \cite{Tenorio:2024jgc}
\begin{equation}
\rho\left(\lambda\right) = \prod_{k=1}^{D} \frac{1}{\delta\lambda[k]},
\end{equation}
where $k$ is the running index over the search parameters, up to $D=6$ in this case.

In order to sample the templates, we perform a coordinate transformation from the physical parameter space $\lambda$ to a uniform template space $\xi$, in which the template density is constant. This is achieved by applying the transformation
\begin{equation}
d\lambda \rho(\lambda) = d\xi \quad \Leftrightarrow \quad \rho(\xi) = 1.
\end{equation}
which we use to derivate the following relations between the coordinates in the physical and uniform parameter space
\begin{equation}
\begin{aligned}
    \xi[0] =& \frac{C^{1/6}}{6} f_0^6 \\
    \xi[1] =& C^{1/6} \alpha \\
    \xi[2] =& C^{1/6} \sin\delta \\
    \xi[3] =& \frac{C^{1/6}}{3} a_{\mathrm{p}}^3 \\
    \xi[4] =& \frac{C^{1/6}}{4} \Omega^{4}\\
    \xi[5] =& C^{1/6} \phi_{\mathrm{b}}
\end{aligned}\,.
\label{eq:unif_transform}
\end{equation}
with $C = \Tsft^{6} T_{\mathrm{obs}} (v/c)^2$ (more details can be found in~\cite{Tenorio:2024jgc}).

In this transformed space, templates are generated by uniform random sampling over the search region. These samples are then mapped back to the physical parameter space $\lambda$ using the inverse transformation as described in~\cite{Tenorio:2024jgc}. As a result, the sampling in $\lambda$ automatically follows the required non-uniform density $\rho(\lambda)$.

The number of templates in a volume region $\Delta\xi$ in the uniform space is simply
\begin{equation}
\mathcal{N}(\Delta\lambda) = \int_{\Delta\xi} d\xi = \prod_{k=1}^{D} \Delta\xi[k].
\label{eq:Ntotal}
\end{equation}

To construct the boxes, the algorithm used is as follows:

We begin by estimating the number of sky templates required due to the Doppler modulation induced by Earth's motion. This number depends on the frequency and the resolution of the sky grid, and is given by
\begin{equation}
\mathcal{N}_{\text{sky}} = \int_{\text{sky}} \frac{1}{\delta \text{sky}} = 4\pi \left(f_0 T_{\text{SFT}} \frac{v}{c} \right)^2.
\end{equation}
We then divide the sky into smaller patches so that each sky patch contains a fixed number of templates. We chose for each sky patch a small number of templates, $\approx9$, in order to ensure that all templates in one box were spatially close to each other. Hence, the total number of sky boxes is
\begin{equation}
B_{\text{sky}} = \frac{\mathcal{N}_{\text{sky}}}{9}.
\end{equation}

We split both sky coordinate parameters into the same number of subdivisions,
\begin{equation}
    B_{\alpha,\sin\delta} = \text{round}\left(\sqrt{B_{\text{sky}}}\right),
    \label{eq:balphasindelta}
\end{equation}
and then calculate the number of templates assigned to each sky box
\begin{equation}
    \mathcal{N}_{\text{per sky box}} = \frac{\mathcal{N}_{\text{total}}\mathcal{O}}{B_{\alpha,\sin\delta}^2}.
\end{equation}
where $\mathcal{O}$ is the oversampling factor for a given frequency band and $\mathcal{N}_{\text{total}}$ the total amount of templates in the frequency band, computed using Eq.~\eqref{eq:Ntotal}. The amount $B_{\alpha,\sin\delta}$ per band is increased in discrete steps according to the increasing frequency (see Eq.~\ref{eq:balphasindelta}). We start at 100~Hz with 12$^2$ sky boxes ending with 42$^2$ at 350~Hz.

We also tune the number of subdivisions in the frequency and binary parameters so that each box contains approximately $10^8$ templates. This limit was enforced by the maximum number of templates that we were able to process at the same time in parallel in a modern NVIDIA A100 Tensor Core GPU as described in~\cite{Tenorio:2024jgc}. Maximizing the number of templates to be processed for the efficient use of GPU resources.

\begin{figure}[ht]
    \includegraphics[width=0.5\textwidth]{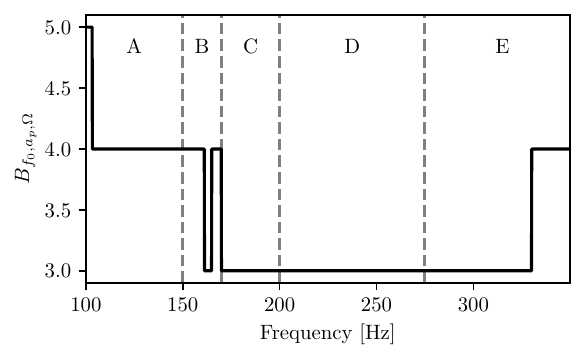}
    \caption{
    Number of subdivisions in the intrinsic frequency $f_0$, projected semi-major axis $a_p$, and orbital frequency $\Omega$ dimensions per 
    0.125~Hz frequency band. The lettered regions (A-E) correspond to the frequency intervals defined in Table~\ref{table:frequency_regions}. 
    Four major transitions can be observed, with different oversampling factors. The first two drops in the number of subdivisions (near 100~Hz and between 
    170-200~Hz) are due 
    to the enforcement of a minimum total number of $10^4$ boxes. The drop at 170~Hz is due to a change in oversampling. At 174~Hz there is another increase because the amount of templates per box is above $1.5\times10^8$, hence, another subdivision is added to round the number closer to $10^8$. The sharp increase beyond 300~Hz is 
    instead driven by the $~10^8$ templates-per-box criterion.}
    \label{fig:bin_param_boxes}

\end{figure}

Hence, we partition the following three parameters: frequency $f_{0}$, projected semi-major axis $a_{p}$, and orbital frequency $\Omega$, into an equal number of subdivisions ($B_{f_0} = B_{a_p} = B_{\Omega} \equiv B_{f_0,a_p,\Omega}$). The quantity $B_{f_0,a_p,\Omega}$ per frequency band is illustrated in Fig.~\ref{fig:bin_param_boxes}. Additionally, orbital phase parameter $\phi_{\mathrm{b}}$ is not subdivided, it is sampled over its full physical range within each box.
The total number of boxes is
\begin{equation}
B_{\text{total}} = B_{\alpha,\sin\delta}^2 B_{f_0,a_p,\Omega}^3,
\end{equation}
and number of templates per box is 
\begin{equation}
\mathcal{N}_{\text{box}} = \frac{\mathcal{N}_{\text{total}}\mathcal{O}}{B_{\text{total}}} = \frac{\mathcal{N}_{\text{total}}\mathcal{O}}{B_{\alpha,\sin\delta}^2 B_{f_0,a_p,\Omega}^3}.
\label{eq:nbox}
\end{equation}
As a consequence, the division of these three parameters is done according to
\begin{equation}
B_{f_0,a_p,\Omega} = \text{round} \left( \sqrt[3]{ \frac{\mathcal{N}_{\text{total}}\mathcal{O}}{\mathcal{N}_{\text{box}} B_{\alpha,\sin\delta}^2} } \right),
\end{equation}
per each 0.125~Hz band, where $\mathcal{N}_{\text{box}}=10^8$.

\begin{figure}[ht]
    \includegraphics[width=0.5\textwidth]{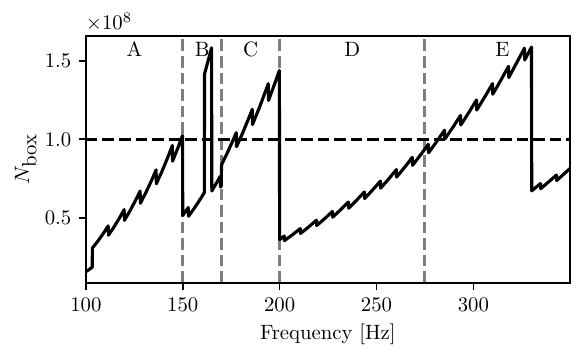}
    \caption{
    Number of templates per box $\bar{\mathcal{N}}_\text{box}$ as a function of frequency. The white dashed line marks the target threshold of $10^8$ 
    templates per box. The five lettered regions (A-E) match the frequency intervals defined in Table~\ref{table:frequency_regions}. The 
    sawtooth pattern arises due to rounding of box counts and the adaptive balancing of box size to the minimum amount of boxes and templates 
    per box criteria as shown in Fig.~\ref{fig:bin_param_boxes}.}
    \label{fig:templates_per_box}
\end{figure}

We ensured that boxes were sufficiently small, in order to avoid that possible contaminations in one of them affected a large region in parameter space, by defining a minimum of $10^4$ boxes per frequency band: if the resulting $\mathcal{N}_{\text{box}}$ from Eq.~\eqref{eq:nbox} is less than $10^4$, we further subdivided the frequency and binary parameter space by incrementing $B_{f_0,a_p,\Omega}$ according to
\begin{equation}
B_{f_{0},a_{p},\Omega}=\left\lceil \sqrt[3]{\frac{B_{\text{min}}}{B_{\alpha,\sin\delta}^{2}}}\right\rceil .
\end{equation}
The total amount of boxes per frequency band is illustrated in Fig.~\ref{fig:total_boxes}, and the number of templates per box in Fig.~\ref{fig:templates_per_box}.

\section{\label{sec:app_sensitivity} Fitting Sensitivity Estimation}

In this appendix, we describe the procedure used to estimate the search sensitivity shown in Sec.~\ref{sec:sensitivity}. Our goal is to quantify the amplitude of CW signals that the search could detect with 95\% efficiency across different frequency bands. This is done by performing a large number of signal injection studies and analyzing their recovery.

After separating detected and non-detected injections, we construct efficiency-versus-depth curves by using six different sensitivity depths. For each depth $\mathcal{D}$, the detection efficiency $E$ is defined as the fraction of successfully detected injections out of the total $N$ performed (500 for each sensitivity depth). Its statistical uncertainty is assigned as
\begin{equation}
\delta E = \sqrt{ \frac{E(1 - E)}{N} },
\end{equation}
the standard error of a proportion when counting successes in a set of independent trials.

To extract the sensitivity depth corresponding to 95\% detection efficiency ($\mathcal{D}^{95\%}$), we fit a sigmoid function to the efficiency-versus-depth data. The fit is performed using the \texttt{curve\_fit} function from \texttt{SciPy}~\cite{Virtanen2020}, with the form
\begin{equation}
S(\mathcal{D}; a, b) = \frac{1}{1 + \exp\left( \frac{\mathcal{D} - b}{a} \right)},
\end{equation}
where $a$ and $b$ are shape parameters. The 95\% sensitivity depth is given by the value of $\mathcal{D}$ for which $S(\mathcal{D}; a, b) = 0.95$.

The uncertainty in the sigmoidal fit is propagated from the covariance matrix $C$ as
\begin{equation}
\sigma_{S}(\mathcal{D}) = \sqrt{
\left( \frac{\partial S}{\partial a} \right)^2 C_{aa} +
\left( \frac{\partial S}{\partial b} \right)^2 C_{bb} +
2 \frac{\partial S}{\partial a} \frac{\partial S}{\partial b} C_{ab}
},
\end{equation}
and the error in the $\mathcal{D}^{95\%}$ is done by bisecting the curves defined by $S(\mathcal{D}; a, b)\pm\sigma_{S}(\mathcal{D})$ with $E=0.95$.

\begin{figure}
\includegraphics[width=0.5\textwidth]{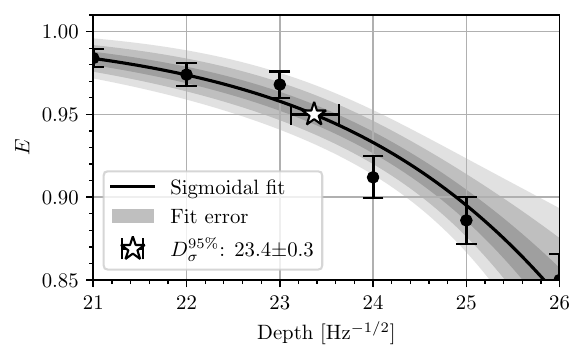}
\caption{Example of $95\%$ sensitivity depth interpolation at 105.75~Hz (Region A, oversampling = 4.0, threshold $\Psi_s > 7.0$). Circles indicate injection results, the black line shows the sigmoidal fit, and shaded bands represent 1, 2, and 3$\sigma$ confidence intervals. The star 
denotes the estimated $\mathcal{D}^{95\%}$ with its $\pm\sigma$ uncertainty.}
\label{fig:eff_example}
\end{figure}

In Fig.~\ref{fig:eff_example}, we show a representative example of the efficiency curve fit used to estimate the $95\%$ sensitivity depth, $\mathcal{D}^{95\%}$, and its error. The $95\%$ depth and its $\pm\sigma$ uncertainty are extracted from the fit and indicated with a star.

\section{\label{sec:vetoed_bands}Vetoed frequency ranges}

The total contaminated frequency range surrounding a line is given by
\begin{equation}
f_{\text{contaminated}} = f_{\text{line}} \pm \left(\Delta f_{\text{line}} + f_{\text{wings}}\right),
\label{eq:contamination_range}
\end{equation}
where $\Delta f_{\text{line}}$ is the width towards each side of the line (left width towards lower frequencies and right width towards higher ones), and $f_{\text{wings}}$ accounts for the maximum Doppler excursion, computed as
\begin{equation}
f_{\text{wings}} = f_{\text{line}} \left( \frac{v}{c} + a_p \Omega \right).
\end{equation}

We provide in this appendix Table~\ref{tab:vetoed_freq_ranges}, which contains a list of the frequency ranges which have been vetoed as described in Sec.~\ref{sec:post_processing}. The contamination range caused by each line has been calculated using Eq.~\eqref{eq:contamination_range}. Outliers from the search in these ranges have been vetoed and have not been followed up in the stage discussed in Sec.~\ref{sec:follow_up}.

\begin{table*}[h!]
\centering
\footnotesize
\begin{tabular}{c@{\hspace{1em}}c@{\hspace{1em}}c@{\hspace{1em}}c@{\hspace{1em}}c@{\hspace{1em}}c}
\toprule
\multicolumn{6}{c}{Vetoed Frequency Ranges (Start, End) [Hz]} \\
\midrule
(100.000, 100.026) & (100.649, 100.702) & (100.860, 100.913) & (101.302, 101.355) & (101.574, 101.699) & (101.969, 102.178) \\
(102.643, 102.696) & (102.963, 103.131) & (103.639, 103.693) & (104.186, 104.313) & (104.624, 104.697) & (105.294, 105.348) \\
(105.632, 105.687) & (106.284, 106.339) & (106.402, 106.457) & (106.629, 106.684) & (107.114, 107.170) & (107.511, 107.566) \\
(107.625, 107.681) & (107.945, 108.001) & (108.619, 108.678) & (109.606, 109.784) & (109.972, 110.028) & (110.615, 110.672) \\
(110.836, 110.893) & (111.081, 111.138) & (111.267, 111.324) & (111.611, 111.677) & (111.944, 112.002) & (112.608, 112.666) \\
(112.927, 112.986) & (113.052, 113.111) & (113.604, 113.796) & (114.161, 114.220) & (114.588, 114.667) & (115.269, 115.329) \\
(115.597, 115.657) & (116.249, 116.309) & (116.378, 116.438) & (116.594, 116.654) & (116.969, 123.091) & (123.134, 123.279) \\
(123.570, 123.634) & (124.136, 124.200) & (124.552, 124.638) & (125.244, 125.309) & (125.563, 125.636) & (126.213, 126.418) \\
(126.559, 126.625) & (127.461, 127.622) & (127.874, 127.940) & (128.552, 128.636) & (129.534, 129.624) & (129.678, 129.745) \\
(129.934, 130.001) & (130.545, 130.613) & (130.786, 130.854) & (130.918, 130.985) & (131.195, 131.263) & (131.542, 131.610) \\
(131.895, 131.963) & (132.538, 132.924) & (133.003, 133.072) & (133.297, 133.366) & (133.535, 133.604) & (134.111, 134.181) \\
(134.516, 134.610) & (135.220, 135.289) & (135.528, 135.598) & (136.177, 136.247) & (136.328, 136.398) & (136.525, 136.595) \\
(137.436, 137.592) & (137.838, 137.909) & (138.518, 138.616) & (139.498, 139.596) & (139.653, 139.725) & (140.511, 140.583) \\
(140.761, 140.834) & (141.159, 141.232) & (141.507, 141.580) & (141.870, 141.943) & (142.078, 142.245) & (142.504, 142.577) \\
(142.820, 143.052) & (143.500, 143.574) & (144.087, 144.161) & (144.405, 144.581) & (145.195, 145.270) & (146.141, 146.217) \\
(146.303, 146.575) & (147.412, 147.488) & (147.802, 147.878) & (148.520, 148.597) & (149.463, 149.566) & (149.628, 149.706) \\
(149.809, 149.886) & (150.737, 150.814) & (151.123, 151.201) & (151.550, 151.728) & (151.845, 151.923) & (152.784, 152.863) \\
(152.953, 153.032) & (153.475, 153.555) & (154.062, 154.141) & (154.445, 154.552) & (155.170, 155.250) & (155.513, 155.594) \\
(156.105, 156.186) & (156.279, 156.359) & (157.387, 157.468) & (157.766, 157.847) & (158.495, 158.577) & (159.427, 159.686) \\
(160.451, 160.535) & (160.712, 160.795) & (161.021, 161.211) & (161.820, 161.904) & (162.748, 162.832) & (162.929, 163.013) \\
(164.037, 164.122) & (164.409, 164.523) & (165.145, 165.231) & (165.757, 165.954) & (166.070, 166.155) & (166.254, 166.339) \\
(166.621, 166.707) & (167.362, 167.515) & (167.730, 167.817) & (168.470, 168.557) & (169.391, 169.666) & (169.984, 170.071) \\
(170.493, 170.775) & (171.052, 171.140) & (171.796, 171.884) & (172.712, 172.801) & (172.904, 172.993) & (174.012, 174.102) \\
(174.373, 174.494) & (175.121, 175.211) & (176.034, 176.125) & (176.229, 176.320) & (176.954, 183.046) & (183.988, 184.082) \\
(184.337, 184.465) & (185.096, 185.191) & (185.998, 186.094) & (186.204, 186.300) & (187.313, 187.409) & (187.659, 187.755) \\
(188.356, 188.518) & (188.838, 188.935) & (189.319, 189.660) & (190.638, 190.736) & (190.980, 191.078) & (191.746, 191.845) \\
(192.641, 192.740) & (192.854, 192.954) & (193.963, 194.063) & (194.302, 194.436) & (195.071, 195.172) & (195.332, 195.434) \\
(195.962, 196.063) & (196.179, 196.280) & (197.288, 197.389) & (197.623, 197.725) & (198.396, 198.498) & (198.909, 199.143) \\
(199.284, 199.421) & (199.505, 199.607) & (199.945, 200.049) & (200.613, 200.716) & (200.944, 201.048) & (201.721, 201.825) \\
(202.605, 202.709) & (202.830, 202.934) & (203.938, 204.043) & (204.266, 204.407) & (205.046, 205.152) & (205.926, 206.032) \\
(206.155, 206.261) & (207.263, 207.370) & (207.587, 207.694) & (208.371, 208.626) & (209.248, 209.588) & (209.733, 209.841) \\
(210.588, 210.697) & (210.909, 211.017) & (211.697, 211.805) & (212.569, 212.679) & (212.805, 212.914) & (213.913, 214.023) \\
(214.230, 214.378) & (215.022, 215.132) & (215.891, 216.002) & (216.130, 216.241) & (217.238, 217.350) & (217.551, 217.663) \\
(217.853, 218.109) & (218.347, 218.459) & (219.212, 219.363) & (219.455, 219.568) & (220.563, 220.677) & (220.873, 220.986) \\
(221.672, 221.786) & (222.533, 222.648) & (222.780, 222.895) & (223.889, 224.004) & (224.194, 224.349) & (224.997, 225.113) \\
(225.855, 225.971) & (226.105, 226.222) & (227.214, 227.633) & (228.322, 228.439) & (229.176, 229.335) & (229.430, 229.548) \\
(229.907, 230.025) & (230.539, 230.657) & (230.837, 230.956) & (231.647, 231.766) & (232.498, 232.617) & (232.755, 232.875) \\
(233.864, 233.984) & (234.158, 234.320) & (234.972, 235.093) & (235.819, 235.940) & (236.080, 236.202) & (236.796, 243.061) \\
(243.839, 243.964) & (244.123, 244.291) & (244.947, 245.073) & (245.783, 245.910) & (246.056, 246.182) & (246.268, 246.559) \\
(247.164, 247.291) & (247.444, 247.571) & (248.272, 248.400) & (249.105, 249.277) & (249.381, 249.509) & (250.489, 250.618) \\
(250.765, 250.894) & (251.598, 251.727) & (252.426, 252.556) & (252.706, 252.836) & (253.814, 253.945) & (254.087, 254.262) \\
(254.923, 255.054) & (255.740, 256.163) & (257.139, 257.272) & (257.408, 257.540) & (258.248, 258.380) & (259.069, 259.248) \\
(259.356, 259.489) & (260.464, 260.598) & (260.729, 260.864) & (261.573, 261.707) & (262.390, 262.525) & (262.681, 262.816) \\
(263.789, 263.925) & (264.051, 264.233) & (264.898, 265.034) & (265.212, 265.525) & (265.712, 265.848) & (266.006, 266.143) \\
(267.115, 267.510) & (268.223, 268.361) & (269.033, 269.219) & (269.331, 269.470) & (269.656, 269.795) & (269.956, 270.095) \\
(270.440, 270.579) & (270.694, 270.833) & (271.548, 271.688) & (272.354, 272.494) & (272.656, 272.797) & (273.765, 273.905) \\
(274.015, 274.204) & (274.684, 275.014) & (275.676, 275.817) & (275.981, 276.123) & (277.090, 277.232) & (277.336, 277.479) \\
(278.198, 278.341) & (278.997, 279.190) & (279.307, 279.450) & (280.415, 280.559) & (280.658, 280.802) & (281.523, 281.668) \\
(282.319, 282.464) & (282.632, 282.777) & (283.208, 283.513) & (283.740, 284.491) & (284.848, 284.995) & (285.640, 285.787) \\
(285.957, 286.104) & (287.065, 287.213) & (287.301, 287.448) & (288.173, 288.321) & (288.961, 289.161) & (289.282, 289.430) \\
(290.390, 290.771) & (291.498, 291.648) & (292.283, 292.433) & (292.607, 292.757) & (293.627, 294.146) & (294.824, 294.975) \\
(295.604, 295.756) & (295.932, 296.084) & (296.923, 303.522) & (303.690, 303.846) & (303.908, 304.117) & (304.799, 304.955) \\
(305.568, 305.725) & (305.852, 306.568) & (307.016, 307.690) & (308.124, 308.282) & (308.890, 309.103) & (309.232, 309.391) \\
(310.341, 310.710) & (311.449, 311.609) & (312.211, 312.372) & (312.557, 312.940) & (313.666, 314.088) & (314.637, 315.474) \\
(315.533, 315.695) & (315.882, 316.045) & (316.991, 317.356) & (318.099, 318.263) & (318.854, 319.074) & (319.207, 319.371) \\
(320.316, 320.480) & (320.515, 320.679) & (321.424, 321.589) & (322.043, 322.423) & (322.533, 322.698) & (323.641, 324.060) \\
(324.749, 324.916) & (325.497, 325.664) & (325.858, 326.025) & (326.966, 327.326) & (328.074, 328.243) & (328.818, 329.045) \\
(329.183, 329.352) & (329.580, 329.750) & (329.880, 330.049) & (330.291, 330.649) & (331.399, 331.907) & (332.140, 332.310) \\
(332.508, 332.679) & (333.616, 334.031) & (334.725, 334.896) & (335.461, 335.633) & (335.833, 336.005) & (336.941, 337.295) \\
(338.050, 338.223) & (338.782, 339.016) & (339.158, 339.332) & & &  \\
\bottomrule
\end{tabular}
\caption{Vetoed frequency ranges due to known instrumental artifacts.}
\label{tab:vetoed_freq_ranges}
\end{table*}

\section{\label{sec:followupappendix} Outlier follow-up setup}

\subsection{First-step follow-up setup}

For the follow-up first stage, we employ 300 walkers with 300 steps per walker. This setup is not expected to achieve full convergence of the MCMC chains. Nonetheless, it is sufficient for our purpose of identifying promising candidates. Better convergence of the MCMC chains is expected to take place if further stages in case a CW signal is detected.

We also add two additional frequency-evolution parameters to be searched over. The first one is the orbital eccentricity $e$. Due to considering non-zero eccentricities, the full frequency-evolution equation for the CW signal is~\cite{Covas:2019jqa}
\begin{align}
f\left(t;\lambda\right) & =f_{0}\bigg\{1+\frac{\vec{v}\left(t\right)}{c}\cdot\hat{n}-a_{p}\Omega\bigg(\cos\left[\Omega\left(t-t_{\text{asc}}\right)\right]\nonumber \\
 & +e\cos\left[\omega+2\Omega\left(t-t_{\text{asc}}\right)\right]\bigg)\bigg\}.
\end{align}
The addition of eccentricity necessitates also the inclusion of the argument of periapsis $\omega$. Furthermore, the first stage is done with the time of periastron passage $t_p$ instead of the time of ascending node $t_{\textrm{asc}}$, since this is the search parameter implemented in the MCMC using \texttt{PyFstat} package~\cite{PhysRevD.97.103020, Keitel2021, pyfstatzenodo, Wette:2020air}. These two are related through
\begin{equation}
t_{\text{asc}}=t_{p}-\frac{\omega}{\Omega},
\end{equation}
where the orbital phase $\phi_{\mathrm{b}}$ can be calculated as
\begin{equation}
\phi_{\mathrm{b}}=\Omega t_{\text{asc}}.
\end{equation}

The priors used for this stage are uniform over intervals centered on the candidate parameters. The size of these prior regions is determined by the resolution Eq.~\eqref{eq:resolutions}, \eqref{eq:resolutions_sky}. Specifically, the frequency $f_0$ is taken within $\pm 1.5$ resolution intervals, the sky position ($\alpha$, $\delta$) and the orbital period $P$ within $\pm 3$ resolution intervals, where the orbital period resolution is given by
\begin{equation}
    \delta P = \frac{P^2\delta\Omega}{2\pi}
\end{equation}
The projected semi-major axis $a_p$ prior is within $\pm 1$ resolution interval, the orbital phase reference time $t_p$ is allowed to vary over the full orbital period, while the argument of periapsis $\omega$ spans the full range $[0, 2\pi]$. The eccentricity $e$ is drawn from the range $[0, e^\text{max}]$, where $e^\text{max}$ is the maximum value to which the search is sensitive, given by Eq.~\eqref{eq:eccentricty}. When the resulting prior boundaries extend beyond the physical limits of a parameter, such as $\delta$ exceeding $\pi/2$, they are truncated to ensure that no samples fall outside.

\subsection{Second-stage follow-up setup}

In the second stage, both the coherence time and the priors are modified. The higher coherence time makes the parameter-space ``peak'' of the signal higher and narrower. For that reason, the priors are also modified with smaller intervals. One additional parameter, the possible spindown or spin-up of the signal, $|\dot{f}_0|$, is also added to this stage.

As shown in \cite{PhysRevD.110.124049}, the coherence time for stage $j$ is given by
\begin{equation}
T_{\text{coh}}^{(j)} = \mathcal{N}_*^{j/D} T_{\text{coh}}^{(0)},
\end{equation}
where we take $\mathcal{N}_{*}=10^3$ as in~\cite{PhysRevD.110.124049}, $D$ is the number of resolvable parameters, and $T_{\textrm{coh}}^{(0)}$ is the coherence time used in the first stage (12 hours). For the second stage, the additional parameter of the frequency spindown $\dot{f}_0$ is introduced, increasing the number of dimensions to $D=9$. This gives a second-stage coherence time of approximately 26 hours, corresponding to $N_\textrm{seg}=219$ segments (down from $N_\textrm{seg}=474$ in the first stage).

In the second stage, the priors are Gaussian distributions centered on the candidate values recovered from the first stage, with narrower widths than those used previously. These gaussian priors allow for the exploration beyond the expected location if the true signal lies slightly further away. Specifically, the frequency $f_0$ has a Gaussian width of $\delta f_0 / 6$, right ascension $\alpha$ a width of $\delta\alpha / 3$, and declination $\delta$ a width of $\delta\delta / 3$. The projected semi-major axis $a_p$ is assigned a width of $\delta a_p / 9$, the orbital period $P$ a width of $\delta P / 3$, the time of ascending node $t_p$ a width of $5\delta t_p / 3$, the eccentricity $e$ a width of $\delta e / 9$, and the argument of periapsis $\omega$ a width of $\delta \omega / 9$. The spindown $|\dot{f}_0|$ is instead given a uniform prior between minus the maximum and the maximum spindown for which the search remains sensitive, given in Eq.~\eqref{eq:spindown_limit}, allowing both positive (spin-up) and negative (spindown) values. When the Gaussian support extends beyond physical limits, such as when the center $\pm$ ten times the width falls outside the allowed physical or search ranges, the prior is replaced with a uniform distribution covering only the physically valid region.

These prior widths ensure that roughly 99.7\% ($3\sigma$) of the initial walker distribution lies within one-third of the prior volume used in the first stage.
The resolution formulas for the additional parameters introduced in this stage are
\begin{align}
\delta t_{\text{asc}} & =\frac{1}{T_{\text{SFT}}f_{0}a_{p}\Omega^{2}},\\
\delta e & =\frac{1}{T_{\text{SFT}}f_{0}a_{p}\Omega},\\
\delta\omega & =\frac{1}{T_{\text{SFT}}f_{0}a_{p}\Omega e}.
\end{align}
The sampler configuration is kept consistent with the first stage, using the \texttt{ptemcee} algorithm with 300 walkers and 300 steps per walker.

\bibliography{references}

\clearpage
\iftoggle{endauthorlist}{
 %
 %
 \let\author\myauthor
 \let\affiliation\myaffiliation
 \let\maketitle\mymaketitle
 \pacs{}

\input{LSC-Virgo-KAGRA-Authors-Feb-2025-prd}
 \newpage
 \maketitle
}

\end{document}

%% file: LSC-Virgo-KAGRA-Authors-Feb-2025-prd.tex
%

\author{A.~G.~Abac\,\orcidlink{0000-0003-4786-2698}}
\affiliation{Max Planck Institute for Gravitational Physics (Albert Einstein Institute), D-14476 Potsdam, Germany}
\author{I.~Abouelfettouh}
\affiliation{LIGO Hanford Observatory, Richland, WA 99352, USA}
\author{F.~Acernese}
\affiliation{Dipartimento di Farmacia, Universit\`a di Salerno, I-84084 Fisciano, Salerno, Italy}
\affiliation{INFN, Sezione di Napoli, I-80126 Napoli, Italy}
\author{K.~Ackley\,\orcidlink{0000-0002-8648-0767}}
\affiliation{University of Warwick, Coventry CV4 7AL, United Kingdom}
\author{C.~Adamcewicz\,\orcidlink{0000-0001-5525-6255}}
\affiliation{OzGrav, School of Physics \& Astronomy, Monash University, Clayton 3800, Victoria, Australia}
\author{S.~Adhicary\,\orcidlink{0009-0004-2101-5428}}
\affiliation{The Pennsylvania State University, University Park, PA 16802, USA}
\author{D.~Adhikari}
\affiliation{Max Planck Institute for Gravitational Physics (Albert Einstein Institute), D-30167 Hannover, Germany}
\affiliation{Leibniz Universit\"{a}t Hannover, D-30167 Hannover, Germany}
\author{N.~Adhikari\,\orcidlink{0000-0002-4559-8427}}
\affiliation{University of Wisconsin-Milwaukee, Milwaukee, WI 53201, USA}
\author{R.~X.~Adhikari\,\orcidlink{0000-0002-5731-5076}}
\affiliation{LIGO Laboratory, California Institute of Technology, Pasadena, CA 91125, USA}
\author{V.~K.~Adkins}
\affiliation{Louisiana State University, Baton Rouge, LA 70803, USA}
% ...

%
%

%% file: P2000488_v28.tex
%
This material is based upon work supported by NSF's LIGO Laboratory, which is a
major facility fully funded by the National Science Foundation.
The authors also gratefully acknowledge the support of
the Science and Technology Facilities Council (STFC) of the
United Kingdom, the Max-Planck-Society (MPS), and the State of
Niedersachsen/Germany for support of the construction of Advanced LIGO 
and construction and operation of the GEO\,600 detector. 
Additional support for Advanced LIGO was provided by the Australian Research Council.
The authors gratefully acknowledge the Italian Istituto Nazionale di Fisica Nucleare (INFN),  
the French Centre National de la Recherche Scientifique (CNRS) and
the Netherlands Organization for Scientific Research (NWO)
for the construction and operation of the Virgo detector
and the creation and support  of the EGO consortium. 
The authors also gratefully acknowledge research support from these agencies as well as by 
the Council of Scientific and Industrial Research of India, 
the Department of Science and Technology, India,
the Science \& Engineering Research Board (SERB), India,
the Ministry of Human Resource Development, India,
the Spanish Agencia Estatal de Investigaci\'on (AEI),
the Spanish Ministerio de Ciencia, Innovaci\'on y Universidades,
the Red Española de Supercomputación - Barcelona Supercomputing Center - Centro Nacional de Supercomputación (BSC-CNS).
the European Union NextGenerationEU/PRTR (PRTR-C17.I1),
the ICSC - CentroNazionale di Ricerca in High Performance Computing, Big Data
and Quantum Computing, funded by the European Union NextGenerationEU,
the Comunitat Auton\`oma de les Illes Balears through the Conselleria d'Educaci\'o i Universitats,
the Conselleria d'Innovaci\'o, Universitats, Ci\`encia i Societat Digital de la Generalitat Valenciana and
the CERCA Programme Generalitat de Catalunya, Spain,
the Polish National Agency for Academic Exchange,
the National Science Centre of Poland and the European Union - European Regional
Development Fund;
the Foundation for Polish Science (FNP),
the Polish Ministry of Science and Higher Education,
the Swiss National Science Foundation (SNSF),
the Russian Science Foundation,
the European Commission,
the European Social Funds (ESF),
the European Regional Development Funds (ERDF),
the Royal Society, 
the Scottish Funding Council, 
the Scottish Universities Physics Alliance, 
the Hungarian Scientific Research Fund (OTKA),
the French Lyon Institute of Origins (LIO),
the Belgian Fonds de la Recherche Scientifique (FRS-FNRS), 
Actions de Recherche Concert\'ees (ARC) and
Fonds Wetenschappelijk Onderzoek - Vlaanderen (FWO), Belgium,
the Paris \^{I}le-de-France Region, 
the National Research, Development and Innovation Office of Hungary (NKFIH), 
the National Research Foundation of Korea,
the Natural Sciences and Engineering Research Council of Canada (NSERC),
the Canadian Foundation for Innovation (CFI),
the Brazilian Ministry of Science, Technology, and Innovations,
the International Center for Theoretical Physics South American Institute for Fundamental Research (ICTP-SAIFR), 
the Research Grants Council of Hong Kong,
the National Natural Science Foundation of China (NSFC),
the Israel Science Foundation (ISF),
the US-Israel Binational Science Fund (BSF),
the Leverhulme Trust, 
the Research Corporation,
the National Science and Technology Council (NSTC), Taiwan,
the United States Department of Energy,
and
the Kavli Foundation.
The authors gratefully acknowledge the support of the NSF, STFC, INFN and CNRS for provision of computational resources.

This work was supported by MEXT,
the JSPS Leading-edge Research Infrastructure Program,
JSPS Grant-in-Aid for Specially Promoted Research 26000005,
JSPS Grant-in-Aid for Scientific Research on Innovative Areas 2402: 24103006,
24103005, and 2905: JP17H06358, JP17H06361 and JP17H06364,
JSPS Core-to-Core Program A.\ Advanced Research Networks,
JSPS Grants-in-Aid for Scientific Research (S) 17H06133 and 20H05639,
JSPS Grant-in-Aid for Transformative Research Areas (A) 20A203: JP20H05854,
the joint research program of the Institute for Cosmic Ray Research,
University of Tokyo,
the National Research Foundation (NRF),
the Computing Infrastructure Project of the Global Science experimental Data hub
Center (GSDC) at KISTI,
the Korea Astronomy and Space Science Institute (KASI),
the Ministry of Science and ICT (MSIT) in Korea,
Academia Sinica (AS),
the AS Grid Center (ASGC) and the National Science and Technology Council (NSTC)
in Taiwan under grants including the Science Vanguard Research Program,
the Advanced Technology Center (ATC) of NAOJ,
and the Mechanical Engineering Center of KEK.

Additional acknowledgements for support of individual authors may be found in the following document: \\
\texttt{https://dcc.ligo.org/LIGO-M2300033/public}.
For the purpose of open access, the authors have applied a Creative Commons Attribution (CC BY)
license to any Author Accepted Manuscript version arising.
We request that citations to this article use 'A. G. Abac {\it et al.} (LIGO-Virgo-KAGRA Collaboration), ...' or similar phrasing, depending on journal convention.

%
%